\documentclass[a4paper,12pt]{article}

\usepackage{amsmath,amssymb}           
\usepackage{amscd}                     
\usepackage{epsfig}                    
\usepackage[matrix,arrow]{xy}          
\usepackage{xspace}                    
\usepackage{stmaryrd}                  
\usepackage{slashed}

\usepackage{jheppub}                   
\makeatletter
\gdef\@fpheader{\ }                    
\gdef\NAT@sort{0}                      
\makeatother

\allowdisplaybreaks[1]                 

\newcommand{\dd}{\mathrm{d}}
\newcommand{\ee}{\mathrm{e}}
\newcommand{\ii}{\mathrm{i}}

\newcommand{\der}{\partial}

\newcommand{\bbZ}{\mathbb{Z}}
\newcommand{\bbR}{\mathbb{R}}
\newcommand{\bbC}{\mathbb{C}}

\DeclareMathOperator{\SU}{\mathit{SU}}
\DeclareMathOperator{\SO}{\mathit{SO}}
\DeclareMathOperator{\USp}{\mathit{USp}}
\DeclareMathOperator{\SL}{\mathit{SL}}
\DeclareMathOperator{\GL}{\mathit{GL}}

\DeclareMathOperator{\Spin}{\mathit{Spin}}
\DeclareMathOperator{\so}{\mathfrak{so}}
\DeclareMathOperator{\su}{\mathfrak{su}}
\DeclareMathOperator{\gl}{\mathfrak{gl}}

\newcommand{\rep}[1]{\mathbf{#1}}
\newcommand{\repp}[2]{(\rep{#1}, \rep{#2})}

\DeclareMathOperator{\vol}{vol}

\newcommand{\ph}[1]{\phantom{#1}}

\newcommand{\Lgen}{L}

\newcommand{\Bgen}[2]{\big\llbracket#1,#2\big\rrbracket}
\newcommand{\BLie}[2]{\left[#1,#2\right]}

\newcommand{\Dgen}{{D}}
\newcommand{\DGVP}{D^{\text{GVP}}}

\DeclareMathOperator{\adj}{ad}
\DeclareMathOperator{\AdS}{AdS}

\newcommand{\volG}{\left|\vol_G\right|}

\newcommand{\proj}[1]{\times_{#1}}

\newcommand{\oadj}{\proj{\text{ad}}}

\DeclareMathOperator{\Edd}{\mathit{E_{d(d)}}}

\DeclareMathOperator{\E7}{\mathit{E}_{7(7)}}

\newcommand{\tA}{{\tilde{A}}}
\newcommand{\tF}{{\tilde{F}}}

\newcommand{\yi}{{\underline{m}}}
\newcommand{\yj}{{\underline{n}}}
\newcommand{\yk}{{\underline{p}}}
\newcommand{\yl}{{\underline{q}}}

\newcommand{\xa}{{\alpha}}
\newcommand{\xb}{{\beta}}

\newcommand{\za}{{\hat{\alpha}}}
\newcommand{\zb}{{\hat{\beta}}}
\newcommand{\zc}{{\hat{\gamma}}}

\newcommand{\beq}{\begin{equation}}
\newcommand{\eeq}{\end{equation}}
\newcommand{\ba}{\begin{array}}
\newcommand{\ea}{\end{array}}

\title{Spheres, generalised parallelisability and consistent
   truncations}

\author[a]{Kanghoon Lee,}
\emailAdd{kanghoon.lee@imperial.ac.uk}
\author[b]{Charles Strickland-Constable}
\emailAdd{charles.strickland.constable@desy.de}
\author[c]{and Daniel Waldram}
\emailAdd{d.waldram@imperial.ac.uk}

\affiliation[a,c]{Department of Physics,
   Imperial College London, \\
   Prince Consort Road, London, SW7 2AZ, UK}

\affiliation[a]{Center for Quantum Spacetime, Sogang University, \\
   Seoul 121-742, Korea}

\affiliation[b]{II. Institut f\"ur Theoretische Physik
   der Universit\"at Hamburg, \\
   Luruper Chaussee 149, D-22761 Hamburg, Germany}

\subheader{\textrm{Imperial/TP/14/DW/0 \\ ZMP-HH/14-3}}

\abstract{We show that generalised geometry gives a unified
   description of maximally supersymmetric consistent truncations of
   ten- and eleven-dimensional supergravity. In
   all cases the reduction manifold admits a ``generalised
   parallelisation'' with a frame algebra with constant
   coefficients. The consistent truncation then arises as a
   generalised version of a conventional Scherk--Schwarz reduction
   with the frame algebra encoding the embedding tensor of the reduced
   theory. The key new result is that all round-sphere $S^d$
   geometries admit such generalised parallelisations with an
   $\SO(d+1)$ frame algebra. Thus we show that the remarkable consistent
   truncations on $S^3$, $S^4$, $S^5$ and $S^7$ are in fact simply 
   generalised Scherk--Schwarz reductions. This description leads
   directly to the standard non-linear scalar-field ansatze and as an
   application we give the full scalar-field ansatz for the type IIB
   truncation on $S^5$.}  

\begin{document}
\maketitle


\section{Introduction}
\label{sec:intro}

Consistent truncations of gravitational theories are few and far
between~\cite{DNP,CGLP}. The classic example is compactification on a
local group manifold $M=G/\Gamma$, where $\Gamma$ is a discrete,
freely-acting subgroup of a Lie group $G$. If the discrete group acts
on the left, the left-invariant vector fields $\hat{e}_a$ define a
global frame so $M$ is parallelisable. Furthermore taking the Lie bracket 
\begin{equation}
\label{eq:Lie-alg}
   \BLie{\hat{e}_a}{\hat{e}_b} = f_{ab}{}^c \hat{e}_c
\end{equation}
the coefficients $f_{ab}{}^c$ are constant. If in addition the
``unimodular'' condition $f_{ab}{}^b=0$ is satisfied then one has a
consistent truncation~\cite{ss}. If the theory is pure metric, the
scalar fields in the truncated theory come from deformations of the
internal metric.
One defines a new global frame
\begin{equation}
   \hat{e}'_a(x) = U_a{}^b(x) \hat{e}_a
\end{equation}
where $U_a{}^b(x)$ depends on the uncompactified coordinates $x$. This
frame defines the vielbein for the transformed metric. By construction
the scalar fields $U_a{}^b(x)$ parameterised a $\GL(d,\bbR)/O(d)$
coset. The truncated  theory is gauged by the group $G$ with the Lie
algebra given by the Lie bracket~\eqref{eq:Lie-alg}.

More generally, as first considered by Scherk and Schwarz~\cite{ss},
any field theory can be reduced on $M$ using left-invariant objects,
and by definition the resulting truncation will be consistent. In
particular, one can consider reductions of heterotic, type II or
eleven-dimensional
supergravity~\cite{KM,CGLP,group-red,dAFT,HR-E1}. Since the 
parallelisation means the tangent space is trivial, $M$ also admits
global spinors and the truncated theories have the same number of
supersymmetries as the original supergravity theory. The structure of
such gauged supergravity theories is very elegantly captured by the
embedding tensor formalism~\cite{embed-lecture}.

In addition to these local group manifold reductions, there is a
famous set of remarkable consistent reductions on
spheres, notably $S^7$~\cite{dWN-S7} and $S^4$~\cite{NVvN} for
eleven-dimensional supergravity, $S^5$ for
type IIB (for which a subsector is known to be
consistent~\cite{CLPST1}), and $S^3$ for the NSNS sector of type II 
supergravity~\cite{CLPST2}. However, generically reductions on coset 
spaces are not consistent and there is ``no known algorithmic
prescription''~\cite{CGLP} for understanding the appearance of these
few special cases.

In this paper we argue for a systematic understanding
of consistent truncations in terms of generalised geometry. In
generalised geometry one considers structures on an generalised tangent
space $E$. In the original formulation~\cite{GCY,Gualtieri} $E\simeq
TM\oplus T^*M$, and the structure on $E$, together with the natural
analogue of the Levi--Civita connection, capture the NSNS degrees of
freedom of type II theories and the bosonic and fermionic equations of
motion~\cite{csw1} (see~\cite{siegel} and also~\cite{JLP} for earlier
geometric reformulations using the closely related Double Field Theory
formalism~\cite{dft}). There are also other versions of generalised
geometry~\cite{chris,PW,baraglia,charlie} with structures and 
connections which capture, for example, the full set of bosonic fields
and equations of motion of type II and eleven-dimensional
supergravity~\cite{csw2,csw3}. The central point for us is that in
each case there is a direct generalised geometric analogue of a local
group manifold, namely a manifold equipped with a global frame
$\{\hat{E}_A\}$ on $E$ such that  
\begin{equation}
\label{eq:embed-def}
   \Lgen_{\hat{E}_A} \hat{E}_B = X_{AB}{}^C \hat{E}_C , 
\end{equation}
where $X_{AB}{}^C$ are constant. By definition $E$ is then trivial and we
say the frame defines a ``generalised parallelisation'' of
$M$~\cite{GMPW}. Since $E$ is trivial, the related generalised spinor
bundle~\cite{chris} is also trivial and hence one also has globally
defined spinors. Thus we expect any truncated theory to have the same number of
supersymmetries as the original supergravity. Just as for the pure
metric case, one can define a ``generalised Scherk--Schwarz''
reduction by defining a rotated generalised frame
\begin{equation}
\label{eq:gen-SS}
   \hat{E}'_A(x) = U_A{}^B(x) \hat{E}_B .
\end{equation}
One is led to conjecture:
\begin{quote}
\emph{Given a generalised parallelisation $\{\hat{E}_A\}$
   satisfying~\eqref{eq:embed-def} there is a consistent truncation on
   $M$ preserving the same number of supersymmetries as the original
   theory with embedding tensor given by $X_{AB}{}^C$ and scalar
   fields encoded by~\eqref{eq:gen-SS}.}
\end{quote}
For compactifications on local group manifolds the conventional global
frame $\{\hat{e}_a\}$ always defines a generalised global frame, and
this conjecture has already, at least implicitly, appeared in the
literature~\cite{KM,HR-E1,GMPW}. In addition, without assuming
a consistent truncation, the relation between the frame algebra and the
embedding tensor of the reduced theory has been
identified~\cite{ABMN,csw2,relax,other-GSS} both in conventional
generalised geometry and in the language of Double Field
Theory~\cite{dft} and its M-theory extensions~\cite{BP-4d}. The
generalised Scherk--Schwarz ansatz~\eqref{eq:gen-SS} is also
in practise used, for the metric components, in the original work on
$S^7$~\cite{dWNW,dWN-S7}, and, recently, this has been extended
to all the flux components~\cite{N-new}. In~\cite{N-embed} the
four-dimensional embedding tensor for conventional Scherk--Schwarz
reductions was also calculated from eleven dimensions using the
``generalised vielbein postulate'' which, as we discuss in the
conclusions, is connected to the algebra~\eqref{eq:embed-def}. 

The key point of this paper is to show that above conjecture also
includes the sphere truncations. In contrast to the case of
conventional geometry where it is a famous result that only $S^1$,
$S^3$ and $S^7$ are parallelisable~\cite{spheres}, we show that,
within an appropriate notion of generalised geometry,
\begin{quote}
\emph{All spheres $S^d$ are generalised parallelisable.}
\end{quote}
Furthermore we show for the round spheres they admit a frame with
constant coefficients $X_{AB}{}^C$ encoding a $\SO(d+1)$ gauging. In
the cases of $S^3$, $S^4$, $S^5$ and $S^7$ this generalised geometry
(or an extension of it) encodes the appropriate ten- or
eleven-dimensional supergravity. In particular we show that the frame
algebra~\eqref{eq:embed-def} reproduces the appropriate embedding
tensor for the $\SO(d+1)$ gauging of the reduced theory, and the
generalised Scherk--Schwarz deformations~\eqref{eq:gen-SS} match the
standard scalar field ansatz for sphere consistent
truncations~\cite{NVvN,NV,CLPST1,CLP}. In the $S^7$ case, we should
note that the tensor components of the parallelising generalised frame 
have recently appeared in~\cite{N-new} building on the seminal work
of~\cite{dWN,dWN-S7}.

The paper is organised as follows. In section~\ref{sec:S-GG} we define
the $\GL^+(d+1,\bbR)$ generalised geometry relevant to the $S^d$
generalised parallelisations. We define the global generalised frame,
show that~\eqref{eq:embed-def} defines an $\so(d+1)$ Lie algebra, and
describe the generalised Scherk--Schwarz reduction of the scalar
fields. Section~\ref{sec:CT} describes how this structure encodes the
classic sphere consistent truncations on $S^3$, $S^4$, $S^5$ and
$S^7$. As an application we derive the general scalar-field ansatz for
the $S^5$ truncation of type IIB. Section~\ref{sec:concl} gives our
conclusions.


\section{Spheres and generalised geometry}
\label{sec:S-GG}


Let us start by showing how the round sphere $S^d$ with a $d$-form
field strength $F$ has a very natural interpretation as a
parallelisation of a particular version of generalised
geometry. This will provide the basic construction for each of our
supergravity examples.


\subsection{The set-up}
\label{sec:set-up}

Consider a theory in $d$ dimensions with metric $g$ and $d$-form field
strength $F=\dd A$, satisfying the equations of motion
\begin{equation}
   R_{mn} = \frac{1}{d-1}F^2 g_{mn} , \qquad
   F = \frac{d-1}{R} \vol_g ,
\end{equation}
where $F^2=\frac{1}{d!}F^{m_1\dots m_d}F_{m_1\dots m_d}$. This admits
a solution with a round sphere $S^d$ metric of radius $R$.

We define various relevant geometrical objects on $S^d$ in
Appendix~\ref{app:sphere}. Here we simply note that, in terms of
constrained coordinates $\delta_{ij}y^iy^j=1$ with $i,j=1,\dots,d+1$,
we can write the metric of radius $R$ on $S^d$ as
\begin{equation}
   \dd s^2 = R^2 \delta_{ij} \dd y^i \dd y^j = R^2 \dd s^2(S^d) .
\end{equation}
There are $d+1$ conformal Killing vectors $k_i$ which satisfy
\begin{equation}
   k_i(y_j) = i_{k_i}\dd y^j = \delta_{ij} - y_iy_j , \qquad
   g^{mn} = R^{-2} \delta^{ij} k_i^m k_j^n , 
\end{equation}
with $\mathcal{L}_{k_i} g = -2 y_i g$. The rotation Killing vectors
can be written as
\begin{equation}
\label{eq:KV-def}
   v_{ij} = R^{-1}\left( y_ik_j - y_jk_i \right) , 
\end{equation}
with the $SO(d+1)$ algebra under the Lie bracket
\begin{equation}
\label{eq:v-alg}
   \BLie{v_{ij}}{v_{kl}}
      = R^{-1}\left(\delta_{ik}v_{lj} - \delta_{il}v_{kj}
          - \delta_{jk}v_{li} + \delta_{jl}v_{ki} \right) .
\end{equation}
%


\subsection{$\GL^+(d+1,\bbR)$ generalised geometry}
\label{sec:GL-gen-geom}

The original formulation of generalised geometry due to Hitchin and
Gualtieri~\cite{GCY,Gualtieri}, considers structures on a
generalised tangent space $E\simeq TM\oplus T^*M$. There is a natural
action of $O(d,d)\times\bbR^+$ on the corresponding frame bundle, and
defining an $O(d)\times O(d)$ sub-structure, or equivalently a
generalised metric $G$, captures the NSNS degrees of freedom of type II
theories. However, this is only one of family of possible generalised
geometries where one considers structures on different generalised
tangent spaces~\cite{chris,PW,baraglia,charlie}. These
capture the bosonic degrees of freedom of the bosonic fields of other
supergravity theories, in particular those of type II and
eleven-dimensional supergravity.

Since the sphere background has a $d$-form field strength it is
natural to consider a generalised geometry with a
$\frac{1}{2}d(d+1)$-dimensional generalised tangent
space,
\begin{equation}
\label{eq:E}
   E \simeq TM \oplus \Lambda^{d-2}T^*M .
\end{equation}
One can write generalised vectors $V=v+\lambda\in E$ or, in components, as
\begin{equation}
   V^M = \begin{pmatrix}
          v^m \\ \lambda_{m_1\dots m_{d-2}}
       \end{pmatrix} .
\end{equation}
As usual $E$ is really defined as an extension
\begin{equation}
\label{eq:twistE}
  0 \longrightarrow \Lambda^{d-2}T^*M \longrightarrow E
      \longrightarrow TM \longrightarrow 0 .
\end{equation}
If locally $F=\dd A$ and $A$ is patched by
\begin{equation}
   A_{(i)} = A_{(j)} + \dd\Lambda_{(ij)} \qquad \text{on $U_i\cap U_j$}
\end{equation}
then the patching of $E$ is given by
\begin{equation}
\label{eq:patch}
   v_{(i)} + \lambda_{(i)}
      = v_{(j)} + \lambda_{(j)} + i_{v_{j}}\dd\Lambda_{(ij)}
\end{equation}
where $v_{(i)}\in TU_i$ and $\lambda_{(i)}\in\Lambda^{n-2}T^*U_i$. This
means that, given a vector $\tilde{v}$, a form $\tilde{\lambda}$,
and a connection $A$ then
\begin{equation}
\label{eq:iso}
   V = \tilde{v} + \tilde{\lambda} + i_{\tilde{v}}A
      = \ee^{A} \tilde{V}
\end{equation}
is a section of $E$, where the last equation is just a definition of
the ``$A$-shift'' operator $\ee^A$. In other words a choice of
connection $A$ defines an isomorphism between sections $\tilde{V}$ of
$TM\oplus\Lambda^{d-2}T^*M$ and sections $V$ of $E$.

Given a pair of sections $V=v+\lambda$ and $W=w+\mu$ the Dorfman or
generalised Lie derivative is just the standard Dorfman
bracket~\cite{GCY,Gualtieri}
\begin{equation}
   \Lgen_V W = \BLie{v}{w} + \mathcal{L}_v\mu - i_w\dd\lambda
\end{equation}
One can also define the corresponding Courant bracket as the
antisymmetrization
\begin{equation}
   \Bgen{V}{W} = \tfrac{1}{2} \left( \Lgen_V W - \Lgen_W V\right) .
\end{equation}

This particular extension of the tangent space gives an interesting
generalised geometry because there is a natural action of
positive determinant transformations $\GL^+(d+1,\bbR)$ on $E$, where
sections transform in the $\frac{1}{2}d(d+1)$-dimensional bivector
representation~\cite{charlie}. (The case of $d=4$ was first considered
in~\cite{DL,chris,BP-4d,BP-alg}.) Concretely, we write the generalised
vector index $M$ as an antisymmetric pair $[\yi\yj]$ of $\GL^+(d+1,\bbR)$
indices, where $\yi,\yj=1,\dots,d+1$,  so that
\begin{equation}
   V^M = V^{\yi\yj}
      = \begin{cases}
            V^{m,d+1} = v^m &\in TM \\
            V^{mn} = \lambda^{mn}
               &\in \Lambda^2TM\otimes \det T^*M
            \end{cases}
\end{equation}
where we are using the isomorphism $\Lambda^2TM\otimes\det T^*M\simeq
\Lambda^{d-2}T^*M$ between bivector densities and $(d-2)$-forms given by
\begin{equation}
    \lambda^{mn}
       = \frac{1}{(d-2)!}\epsilon^{mnp_1\dots p_{d-2}}
          \lambda_{p_1\dots p_{d-2}} ,
\end{equation}
where $\epsilon^{m_1\dots m_d}$ is the totally antisymmetric symbol,
with components taking the values $\pm1$. The $\GL^+(d+1,\bbR)$ Lie
algebra acts as 
\begin{equation}
   \delta V^{\yi\yj} = R^\yi{}_\yk V^{\yk\yj}
      + R^\yj{}_\yk V^{\yi\yk} ,
\end{equation}
and we can parameterise the Lie algebra element as
\begin{equation}
   R^\yi{}_\yj
      = \begin{pmatrix}
         r^m{}_n - \frac{1}{2}r^p{}_p\delta^m{}_n
             + \frac{1}{2}c\delta^m{}_n & a^m \\
         \alpha_n & \frac{1}{2}r^p{}_p + \frac{1}{2}c
      \end{pmatrix} .
\end{equation}
where
\begin{equation}
\label{eq:A-iso}
\begin{aligned}
   a^m &= \frac{1}{(d-1)!} \epsilon^{mp_1\dots p_{d-1}}
      a_{p_1\dots p_{d-1}} && \in TM\otimes\det T^*M
         \simeq\Lambda^{d-1}T^*M , \\
   \alpha_m &= \frac{1}{(d-1)!} \epsilon_{mp_1\dots p_{d-1}}
      \alpha^{p_1\dots p_{d-1}} && \in T^*M\otimes\det TM
         \simeq\Lambda^{d-1}TM .
\end{aligned}
\end{equation}
In terms of $v$ and $\lambda$ we have
\begin{equation}
\label{eq:adj-action}
\begin{aligned}
   \delta v^{ m} &= c v^m + r^m{}_n v^n
      - \tfrac{1}{(d-2)!}\alpha^{mn_1\dots n_{d-2}}
          \lambda_{n_1\dots n_{d-2}} , \\
   \delta \lambda_{m_1\dots m_{d-2}} &= c \lambda_{m_1\dots m_{d-2}}
      - (d-2) r^n{}_{[m_1} \lambda_{|n|m_2\dots m_{d-2}]}
      + v^n a_{nm_1\dots m_{d-2}} ,
\end{aligned}
\end{equation}
and we see that $r^m{}_n$ parameterises the usual $\GL(d,\bbR)$
action on tensors. We see that the corresponding adjoint bundle
$\adj\hat{F}$ decomposes as 
\begin{equation}
   \adj{\hat{F}} \simeq \bbR \oplus (TM\otimes T^*M)
       \oplus \Lambda^{d-1}TM \oplus \Lambda^{d-1}T^*M
\end{equation}
and is indeed $(d+1)^2$-dimensional. Note that $a$ generates the
``$A$-shift'' transformation~\eqref{eq:iso}. Also setting
$c=\tfrac{(d-3)}{(d+1)}r^p{}_p$ generates the $\SL(d+1,\bbR)$
subgroup.

The partial derivative $\der_m$ naturally lives in the dual
generalised vector space $E^*\simeq  T^*M\oplus\Lambda^{d-2}TM$ as
\begin{equation}
   \der_M = \der_{\yi\yj}
      = \begin{cases}
         \der_{m,d+1} = \der_m \\
         \der_{mn} = 0
      \end{cases} . 
\end{equation}
One can write the generalised Lie derivative in $\GL^+(d+1,\bbR)$ form
via the usual formula~\cite{csw2}
\begin{equation}
\label{eq:Lgen}
   (\Lgen_V W)^M= (V\cdot\der)W^M - (\der \oadj V)^M{}_N W^N ,
\end{equation}
where $V\cdot U$ denotes the contraction between elements of
$E$ and $E^*$, while $U \oadj V$ is the projection from $E^*\otimes E$
onto the adjoint representation of Lie algebra
$\gl(d+1,\bbR)$. Concretely we have\footnote{\label{sum-conv}
   Throughout this paper whenever there is a an implied sum over $p$
   antisymmetric indices, as the in $V^MU_M$ in the first line
   of~\eqref{eq:dot-adj}, our conventions are that the sum comes with
   a weight of $1/p!$.}
\begin{equation}
\label{eq:dot-adj}
\begin{aligned}
   V\cdot U = V^MU_M &= \tfrac{1}{2} V^{\yi\yj} U_{\yi\yj} , \\
   (U \oadj V)^\yi{}_\yj
      &= V^{\yi\yk} U_{\yj\yk} - \tfrac14 V^{\yk \yl} U_{\yk \yl} \delta^{\yi}{}_{\yj}.
\end{aligned}
\end{equation}
The form of $\Lgen_V$ given in \eqref{eq:Lgen} naturally extends to an
action on any given $\GL^+(d+1,\bbR)$ representation.

As usual the bosonic degrees of freedom $g$ and $A$, together with an
extra overall scale factor $\Delta$, parameterise a generalised metric
$G_{MN}$. Here $G$ is invariant under an $\SO(d+1)\subset\GL^+(d+1,\bbR)$
subgroup. Concretely, if $V=\ee^\Delta \ee^A\tilde{V}$, and using the
definition~\eqref{eq:dot-adj} of the contraction $V^M U_N$, we
have (cf.~\cite{DL,chris,BP-4d,csw3} and see also~\cite{JS})
\begin{equation}
\label{eq:G}
\begin{aligned}
   G&(V,V) = G_{MN} V^M V^N \\
      &=  g_{mn}\tilde{v}^m\tilde{v}^n
          + \tfrac{1}{(d-2)!} g^{m_1n_1}\dots g^{m_{d-2}n_{d-2}}
              \tilde{\lambda}_{m_1\dots m_{d-2}}
              \tilde{\lambda}_{n_1\dots n_{d-2}} \\
     &= V^T \cdot \ee^{-2\Delta} \begin{pmatrix}
           g_{mn} + \tfrac{1}{(d-2)!}A_m{}^{n_1\dots n_{d-2}}
               A_{nn_1\dots n_{d-2}} & -A_m{}^{n_1\dots n_{d-2}} \\
           -A_n{}^{m_1\dots m_{d-2}}  &
           (d-2)! g^{m_1\dots m_{d-2},n_1\dots n_{d-2}}
        \end{pmatrix} \cdot V
\end{aligned}
\end{equation}
where $g^{m_1\dots m_{d-2},n_1\dots n_{d-2}}$ is short-hand for
$g^{[m_1|n_1|}\dots g^{m_{d-2}]n_{d-2}}$ antisymmetrised separately on
the sets of $m_i$ and $n_i$ indices. The factor $\Delta$ is related to
warped compactifications in supergravity theories~\cite{csw2,csw3} as
we will see. 

Another way to view the generalised metric, and see more explicitly
that it is invariant under $\SO(d+1)$, is to note that we can also
consider generalised tensors that transform in the fundamental
$(d+1)$-dimensional representation of $\GL^+(d+1,\bbR)$. We define a
$(d+1)$-dimensional bundle of weighted vectors and densities, as
in~\cite{csw3} for the case $d=4$,
\begin{equation}
\label{eq:fund-rep}
   W \simeq (\det T^*M)^{1/2} \otimes \left( TM \oplus \Lambda^dTM
      \right) ,
\end{equation}
where sections $K=q+t\in W$ can be labelled as
\begin{equation}
   K^\yi = \begin{cases}
         V^{m} = q^m &\in (\det T^*M)^{1/2} \otimes TM \\
         V^{d+1} = t &\in (\det T^*M)^{-1/2}
      \end{cases} , 
\end{equation}
and we are using the isomorphism $(\det
T^*M)^{1/2}\otimes\Lambda^dTM\simeq (\det T^*M)^{-1/2}$. By
construction $E=\Lambda^2W$. We then have an $\SO(d+1)$ metric given
by 
\begin{equation}
\label{eq:G-fund}
\begin{aligned}
   G&(K,K) = G_{\yi\yj} K^{\yi} K^{\yj} \\
     &= K^T \cdot \frac{\ee^{-\Delta}}{\sqrt{g}} \begin{pmatrix}
           g_{mn} & g_{mn} A^n \\
           g_{np}A^p \;  &
           \det g + g_{pq} A^p A^q
        \end{pmatrix} \cdot K ,
\end{aligned}
\end{equation}
where $A^m$ is the vector-density equivalent to $A_{m_1\dots m_{d-1}}$
defined in~\eqref{eq:A-iso}. One then has
\begin{equation}
   G(V,V) = \tfrac{1}{2}G_{\yi\yk}G_{\yj\yl} V^{\yi\yj} V^{\yk\yl}.
\end{equation}
giving the generalised metric on $E$.

Just as for Einstein gravity we can always introduce a local
orthonormal frame $\{\hat{E}_A\}$ for $G$. Recall that $E$ transforms
as a bivector under $\GL^+(d+1,\bbR)$. Thus the frame also transforms
as a two-form under $\SO(d+1)$ and so is naturally labelled by an
antisymmetric pair of $\SO(d+1)$ vectors indices, and so we write the
basis generalised vectors as $\{\hat{E}_{ij}\}$ with
$i,j=1,\dots,d+1$. By definition, we have the orthonormal condition
\begin{equation}
\label{eq:ortho}
   G(\hat{E}_{ij},\hat{E}_{kl})
      = \delta_{ik}\delta_{jl} - \delta_{il}\delta_{jk} .
\end{equation}
Given the isomorphism~\eqref{eq:E} one can define a sub-class of
orthonormal frames that transform under an $\SO(d)$ subgroup of
$\SO(d+1)$ and can be written in terms of the conventional orthonormal
frame $\hat{e}_a$, and their dual one-forms $e_a$, defined by the
metric $g$. These are called  ``split frames'' in~\cite{csw1,csw2}, and
here are given by
\begin{equation}
\label{eq:split}
   \hat{E}_{ij}
      = \begin{cases}
          \hat{E}_{a,d+1} = \ee^\Delta \left(
             \hat{e}_a + i_{\hat{e}_a}A \right) \\
          \hat{E}_{ab} = \tfrac{1}{(d-2)!} \ee^{\Delta}
               \epsilon_{abc_1\dots c_{d-2}}
               e^{c_1}\wedge\dots \wedge e^{c_{d-2}}
       \end{cases} .
\end{equation}
Note that, as described in~\cite{csw3} in the case of $d=4$, one can
always introduce a corresponding frame $\hat{E}_i$ on $W$ such that
$\hat{E}_{ij}=\hat{E}_i\wedge\hat{E}_j$. For the split
frame, the corresponding (dual) frame $\{E^i\}\in W^*$ is given by
\begin{equation}
   E^i = \begin{cases}
           E^a =  g^{-1/4}\ee^{-\Delta/2} \left(
               e^a - e^a\wedge A \right) \\
           E^{d+1} = g^{-1/4}\ee^{-\Delta/2} \vol_g
        \end{cases} .
\end{equation}
One can then write the generalised metric $G_{\yi\yj}$
in~\eqref{eq:G-fund} in terms of the dual frame $E^i$ as
\begin{equation}
   G_{\yi\yj} = \delta_{ij} E^i_\yi E^j_\yj .
\end{equation}
It is important to note that any local rotation of the frame
\begin{equation}
   \hat{E}'_{ij} = U_i{}^k U_j{}^l \hat{E}_{kl} , 
\end{equation}
where $U\in\SO(d+1)$, gives an equally good generalised orthonormal
frame. Note that $U$ and $-U$ actually generate the same
transformation. Thus, when $d$ is odd, the local group defined
by the generalised metric is actually $\SO(d+1)/\bbZ_2$.


\subsection{Spheres as generalised parallelisable spaces}
\label{sec:sphere-para}

In conventional geometry a parallelisable space is one that admits a
global frame, that is, where each basis vector $\hat{e}_a$ is a
globally defined smooth vector field. Topologically it means that the
tangent space $TM$ is trivial. It is a famous result due to Bott and
Milner and Kervaire~\cite{spheres} that the only parallelisable
spheres are $S^1$, $S^3$ and $S^7$. Here we show, by explicit
construction, that by contrast every sphere $S^d$ is ``generalised
parallelisable''.

Generalised parallelisability means that the $\GL^+(d+1,\bbR)$
generalised vector bundle~\eqref{eq:E} admits a global generalised
frame and hence is trivial. On the sphere with flux $F=\dd A$, we
define the global frame as
\begin{equation}
\label{eq:globalE}
   \hat{E}_{ij}
      = v_{ij} + \sigma_{ij} + i_{v_{ij}} A
\end{equation}
where $v_{ij}$ are the $\SO(d+1)$ Killing vectors on $S^d$ given
in~\eqref{eq:KV-def} and
\begin{equation}
   \sigma_{ij} = * \left( R^2 \dd y_i\wedge\dd y_j \right)
      = \frac{R^{d-2}}{(d-2)!}\epsilon_{ijk_1\dots k_{d-1}}y^{k_1}
          \dd y^{k_2}\wedge\dots\wedge\dd y^{k_{d-1}} ,
\end{equation}
where the functions $y_i$ are the constrained coordinates
$\delta_{ij}y^iy^j=1$. To see that the frame is globally defined note
that
\begin{equation}
\begin{aligned}
   v_{ij} &= 0  && && && \text{when $y_i=y_j=0$} \\
   \dd y_i\wedge \dd y_j &= 0  && && && \text{when $y^2_i+y^2_j=1$}
\end{aligned}
\end{equation}
so, while the vector and form parts can separately vanish, each
combination $\hat{E}_{ij}$ is always non-zero. By construction, they
are globally defined sections of $E$. Furthermore, from~\eqref{eq:G}
we have
\begin{equation}
   G(\hat{E}_{ij},\hat{E}_{kl})
      = v_{ij}\cdot v_{kl}
         + \sigma_{ij}\cdot\sigma_{kl}
      = \delta_{ik}\delta_{jl} - \delta_{il}\delta_{jk}
\end{equation}
where we have used~\eqref{eq:contractions}. We see that the frame is
orthonormal with respect to the generalised metric on the round
sphere. Note the corresponding globally defined dual frame $E^i$ is
given by
\begin{equation}
\label{eq:Ei}
   E^i = g^{-1/4} \left(
      R\dd y^i + y^i \vol_g - R\dd y^i \wedge A \right).
\end{equation}
which is clearly globally defined and non-vanishing since $\dd y_i=0$
when $y_i^2=1$.

We can also calculate the analogue of the Lie bracket algebra of
$\hat{E}_{ij}$ by calculating the generalised Lie derivatives. One
finds
\begin{equation}
\begin{aligned}
   \Lgen_{\hat{E}_{ij}} \hat{E}_{kl}
      &= \BLie{v_{ij}}{v_{kl}}
         + \mathcal{L}_{v_{ij}}\left( \sigma_{kl} + i_{v_{kl}}A \right)
         - i_{v_{kl}}\dd\left(\sigma_{ij} + i_{v_{ij}} A \right) \\
      &= \BLie{v_{ij}}{v_{kl}}
         + \mathcal{L}_{v_{ij}}\sigma_{kl}
         + i_{\BLie{v_{ij}}{v_{kl}}} A
         - i_{v_{kl}} \left( \dd\sigma_{ij} - i_{v_{ij}} F \right) \\
      &= \BLie{v_{ij}}{v_{kl}}
         + \mathcal{L}_{v_{ij}}\sigma_{kl}
         + i_{\BLie{v_{ij}}{v_{kl}}} A ,
\end{aligned}
\end{equation}
where in going from the second to the third line we have used
$F=R^{-1}(d-1)\vol_g$ and the
identity~\eqref{eq:vol-identity}. Thus by~\eqref{eq:v-alg}
and~\eqref{eq:Lie-ders} we have
\begin{equation}
\label{eq:SO-alg}
   L_{\hat{E}_{ij}} \hat{E}_{kl} = \Bgen{\hat{E}_{ij}}{\hat{E}_{kl}}
      = R^{-1} \big(\delta_{ik}\hat{E}_{lj}
          - \delta_{il}\hat{E}_{kj}
          - \delta_{jk}\hat{E}_{li}
          + \delta_{jl}\hat{E}_{ki} \big) .
\end{equation}
We see that the generalised Lie derivative algebra of the frame is
simply the Lie algebra $\so(d+1)$.


\subsection{Generalised $\SL(d+1,\bbR)$ Scherk--Schwarz
   reduction on $S^d$}
\label{sec:gen-SS}

Recall that, given a conventional parallelisable manifold $M$, if the 
Lie bracket algebra of the frame $\hat{e}_a$
\begin{equation}
   \BLie{\hat{e}_a}{\hat{e}_b} = f_{ab}{}^c \hat{e}_c
\end{equation}
has constant $f_{ab}{}^c$ then the parallelisation defines a Lie
algebra and we have a local group manifold: $M$ is either a Lie group
or a discrete, freely-acting quotient of a Lie group. It is well-known that such
spaces admit consistent truncations~\cite{DNP,CGLP}, provided
$f_{ab}{}^b=0$~\cite{ss}. The standard metric is given by a bilinear
on the Lie algebra, for instance the Killing form, so
\begin{equation}
   g^{mn} = \delta^{ab} \hat{e}^m_a \hat{e}^m_a .
\end{equation}
The scalar fields of the truncated theory correspond to a
Scherk--Schwarz~\cite{ss} reduction. One considers $\GL(d,\bbR)$
rotations of the frame that are constant on $M$ (though depend on the
coordinates $x$ in the non-compact space)
\begin{equation}
   \hat{e}'_a = U_a{}^b(x) \hat{e}_b , \qquad
   g^{\prime mn} = H^{ab}(x) \hat{e}^m_a \hat{e}^m_a ,
\end{equation}
where the symmetric matrix $H^{cd}=\delta^{ab}U_a{}^c U_b{}^d$
parameterises the $\GL(d,\bbR)/O(d)$ coset space of deformations.

We have shown that the $S^d$ sphere is actually a direct generalised
geometric analogue of a local group manifold. It admits a globally
defined orthonormal frame, and the generalised Lie derivative of the
frame defines a Lie algebra $\so(d+1)$. Thus it is natural to consider
a generalised Scherk--Schwarz reduction~\eqref{eq:gen-SS}. The new
generalised frame is given by
\begin{equation}
   \hat{E}'_{ij} = U_i{}^k(x)U_j{}^l(x) \hat{E}_{kl}
\end{equation}
where $U_i{}^j(x)$ are $\GL(d+1,\bbR)$ matrices, constant on $M$. The
new inverse generalised metric is then given by\footnote{Note that the
   factor of $\tfrac{1}{2}$ comes from the
   normalisation~\eqref{eq:ortho}.}
\begin{equation}
   G^{\prime MN} = \tfrac{1}{2}T^{ik}T^{jl} \hat{E}^M_{ij} \hat{E}^N_{kl} .
\end{equation}
where we define the symmetric object
$T^{kl}=\delta^{ij}U_i{}^kU_j{}^l$. In what follows we will actually
only need to consider $\SL(d+1,\bbR)$ transformations so we can take $\det
T=1$. Thus $T^{ij}$ parameterises an $\SL(d+1,\bbR)/SO(d+1)$
coset. Inverting~\eqref{eq:G}, we find the general form of the
inverse metric, in terms of component fields $g'$, $A'$ and warp
factor $\Delta'$,
\begin{equation}
   G^{\prime MN}
     = \ee^{2\Delta'}\begin{pmatrix}
           g^{\prime mn}
           & g^{\prime mp} A'_{pn_1\dots n_{d-1}} \\
           g^{\prime np} A'_{pm_1\dots m_{d-1}} \quad
           & (d-2)! g'_{m_1\dots m_{d-2},n_1\dots n_{d-2}}
              + A'_{pm_1\dots m_{d-2}}
                 A^{\prime'p}{}^{n_1\dots n_{d-2}}
        \end{pmatrix} .
\end{equation}
Comparing the two expressions gives
\begin{equation}
\begin{aligned}
   \ee^{2\Delta'}g^{\prime mn}
      &= \tfrac{1}{2}T^{ik}T^{jl} v^m_{ij} v^n_{kl} , \\
   \ee^{2\Delta'}(A'-A){}_{m_1\dots m_{d-1}}
      &= \tfrac{1}{2}T^{ik}T^{jl} v_{ij,[m_1}
         \sigma_{kl,m_2\dots m_{d-1}]} ,
\end{aligned}
\end{equation}
where the index on $v_{ij}$ in the second line is lowered using
$g'_{mn}$ and $A$ is the fixed potential on the original undeformed
$S^d$. Since we are considering $\SL(d+1,\bbR)$ transformations we
have $\det G'=\deg G$, implying
\begin{equation}
   \ee^{2(d+1)\Delta'}(\det g')^{-1+(d-2)} = (\det g)^{-1+(d-2)} .
\end{equation}
The analysis of the metric then follows from that
in~\cite{NVvN,NV}. Using $i_{v_{ij}}\dd
y_k=R^{-1}\left(y_i\delta_{jk}-y_j\delta_{ik}\right)$
and~\eqref{eq:metric-id} we have
\begin{equation}
\begin{aligned}
   \tfrac{1}{2}\left(T^{ik}T^{jl} v^m_{ij} v^n_{kl}\right)
      \left(T^{-1}_{i'j'}\der_ny^{i'}\der_py^{j'}\right)
      &= R^{-2} \left(T^{ik}y_k\right) v^m_{ij}\der_p y^j , \\
      &= R^{-2} \left(T^{ij}y_iy_j\right) \delta^m{}_n .
\end{aligned}
\end{equation}
Hence, using~\eqref{eq:det-id}, we have
\begin{equation}
\label{eq:scalars}
\begin{aligned}
   \dd s^{\prime 2} &= \frac{R^2}{(T^{kl}y_ky_l)^{2/(d-1)}}
         \; T^{-1}_{ij} \dd y^i \dd y^j , \\
   A' &= -\frac{1}{2(T^{kl}y_ky_l)} \frac{R^{d-1}}{(d-2)!}\;
       \epsilon_{i_1\dots i_{d+1}}
       (T^{i_1j}y_j)y^{i_2} \dd y^{i_3}\wedge
          \dots \wedge \dd y^{i_{d+1}} + A , \\
   \ee^{2\Delta'} &= (T^{kl}y_ky_l)^{(d-3)/(d-1)} .
\end{aligned}
\end{equation}
As we will see, for the cases of interest, this exactly agrees with the
standard scalar field ansatz for sphere
consistent truncations~\cite{dWNW,KPW,NVvN,NV,CLPST1,CLP}.


\section{Consistent truncations on spheres}
\label{sec:CT}


We now discuss how the generalised parallelisability of $S^d$ relates
to the classic supergravity sphere solutions: the $S^3$ near-horizon
NS-fivebrane background, $\AdS_7\times S^4$ in eleven-dimensional
supergravity, $\AdS_5\times S^5$ in type IIB, and $\AdS_4\times S^7$
in eleven-dimensional supergravity.

Each of these examples has a corresponding consistent truncation on
the $S^d$ sphere to a seven-, five- or four-dimensional gauged
supergravity theory. This has been shown explicitly for
$S^7$~\cite{dWN-S7}, $S^4$~\cite{NVvN} and $S^3$~\cite{CLPST2} and for
a subsector of $S^5$~\cite{CLPST1,CLP}. We will consider each example
in turn, demonstrating how the generalised geometry encodes the
embedding tensor and the scalar field ansatz for the consistent
truncation. In particular we give the general scalar ansatz for
the $S^5$ case.


\subsection{$S^3$ and $\SO(3,3)$ generalised geometry}
\label{sec:s3}

The solution of type II supergravity corresponding to the near-horizon
limit of parallel NS fivebranes has the form of a three-sphere times a
linear dilaton background $\bbR^{5,1}\times\bbR_t\times S^3$~\cite{CHS}
\begin{equation}
\begin{aligned}
   \dd s^2 &= \dd s^2(\bbR^{5,1}) + \dd t ^2 + R^2\dd s^2(S^3) , \\
   H &= 2R^{-1} \vol_g , \\
   \phi &= - t/R ,
\end{aligned}
\end{equation}
where $R$ is the radius of the three-sphere.

In terms of $\GL^+(4,\bbR)$ generalised geometry on the $S^3$ the
relevant generalised tangent space is now
\begin{equation}
   E \simeq TM \oplus T^*M ,
\end{equation}
and, since for $d=3$ we can simply set $c=0$ in the
algebra~\eqref{eq:adj-action} and restrict to an $\SL(4,\bbR)$ action.
The structure groups can be viewed as
\begin{equation}
   \SL(4,\bbR) \simeq \SO(3,3) , \quad \text{and} \quad
   \SO(4)/\bbZ_2 \simeq \SO(3)\times\SO(3)
\end{equation}
where we have used the fact that for $d$ odd the generalised metric is
preserved by a $\SO(d+1)/\bbZ_2$ group. We see that we have the
original $O(d,d)$ generalised geometry considered by Hitchin and
Gualtieri~\cite{GCY,Gualtieri}.

The $\SO(4)$ generalised frame is simply\footnote{Comparing
   with~\eqref{eq:globalE}, we have identified $B=-A$ to match the
   usual $O(d,d)$ generalised geometry conventions.}
\begin{equation}
   \hat{E}_{ij} = v_{ij} + \sigma_{ij} - i_{v_{ij}} B
\end{equation}
and the algebra~\eqref{eq:SO-alg} is the
$\so(4)\simeq\so(3)\times\so(3)$ Lie algebra. To see this in a basis
that is more conventional for $O(d,d)$ generalised geometry, first
introduce the usual left- and right-invariant vector fields on $S^3$
\begin{equation}
\begin{aligned}
   l_+ &= l_1+\ii l_2
      = R^{-1} \ee^{-\ii\psi} \big[  \der_\theta
         + \ii\csc\theta \der_\phi - \ii\cot\theta \der_\psi  \big] ,
         & &&
   l_3 &= R^{-1} \der_\psi , \\
   r_+ &= r_1 + \ii r_2
      = R^{-1}\ee^{\ii\phi} \big[  \der_\theta
         + \ii\cot\theta \der_\phi  - \ii\csc\theta \der_\psi \big] ,
         & &&
   r_3  &= R^{-1}\der_\phi ,
\end{aligned}
\end{equation}
with the corresponding left- and right-invariant one-forms
\begin{equation}
\begin{aligned}
   \lambda_+
     &= R \ee^{-\ii\psi} \left(\dd\theta + \ii\cos\theta\dd\phi\right) ,
     & &&
   \lambda_3 &= R \left( \dd\psi + \cos\theta\dd\phi \right) , \\
   \rho_+
     &= R \ee^{\ii\phi} \left(\dd\theta + \ii\sin\theta\dd\psi \right) ,
     & &&
   \rho_3 &= R \left( \dd\phi + \cos\theta\dd\psi \right) .
\end{aligned}
\end{equation}
We also chose a gauge
\begin{equation}
   B= 2R \cos\theta \dd\phi\wedge \dd\psi .
\end{equation}
Defining two $\SO(3)$ triplets $\hat{E}^L_{\bar{a}}$ and $\hat{E}^R_a$ as
the anti-self-dual and self-dual combinations of $\hat{E}_{ij}$ we
have
\begin{equation}
\begin{aligned}
   \hat{E}^L_+ &= l_+ - \lambda_+ - i_{l_+}B \\
      &= \ee^{-\ii\psi} \Big[
         \left( R^{-1}\der_\theta - R\dd\theta \right)
         + \ii\csc\theta \left( R^{-1}\der_\phi - R\dd\phi \right)
         - \ii\cot\theta\left( R^{-1}\der_\psi
               + R\dd\psi \right) \Big] , \\
   \hat{E}^L_3 &= l_3 - \lambda_3 - i_{l_3}B
      = R^{-1}\der_\psi - R\dd\psi , \\
   \hat{E}^R_+ &= r_+ + \rho_+ - i_{r_+}B \\
      &= \ee^{\ii\phi} \Big[
         \left( R^{-1}\der_\theta + R\dd\theta \right)
         + \ii\cot\theta \left( R^{-1}\der_\phi - R\dd\phi \right)
         - \ii\csc\theta\left( R^{-1}\der_\psi + R\dd\psi \right)
         \Big] , \\
   \hat{E}^R_3 &= r_3 + \rho_3 - i_{r_3}B
      = R^{-1}\der_\phi + R\dd\phi .
\end{aligned}
\end{equation}
These are the conventional left and right bases for the two $\SO(d)$
groups in generalised geometry (see for example~\cite{csw1} where they
are labelled $\hat{E}^-_{\bar{a}}$ and $\hat{E}^+_a$). They are
orthonormal in the sense that, defining
\begin{equation}
   \hat{E}_A = \begin{pmatrix} \hat{E}^R_a \\
         \hat{E}^L_{\bar{a}} \end{pmatrix} ,
\end{equation}
we have
\begin{equation}
\begin{aligned}
   \eta(\hat{E}_A,\hat{E}_B)
      &= \begin{pmatrix}
            \delta_{ab} & 0 \\
            0 & - \delta_{\bar{a}\bar{b}}
        \end{pmatrix} , \\
   G(\hat{E}_A,\hat{E}_B)
      &= \begin{pmatrix}
            \delta_{ab} & 0 \\
            0 & \delta_{\bar{a}\bar{b}}
        \end{pmatrix} ,
\end{aligned}
\end{equation}
where $\eta$ is the usual $O(3,3)$ metric, that is, if $V=v+\lambda$,
\begin{equation}
   \eta(V,V) = i_v\lambda ,
\end{equation}
and $G$ is the generalised metric~\eqref{eq:G} (with $\Delta=0$). Under
the generalised Lie derivative the algebra reads
\begin{equation}
\label{eq:su2su2}
\begin{aligned}
  \Lgen_{\hat{E}^L_{\bar{a}}}\hat{E}^L_{\bar{b}}
     = \Bgen{\hat{E}^L_{\bar{a}}}{\hat{E}^L_{\bar{b}}}
     &= R^{-1}\epsilon_{\bar{a}\bar{b}\bar{c}}
        \hat{E}^L_{\bar{c}} , \\
  \Lgen_{\hat{E}^R_a}\hat{E}^R_b
     = \Bgen{\hat{E}^R_a}{\hat{E}^R_b}
     &= R^{-1}\epsilon_{abc} \hat{E}^R_c , \\
  \Lgen_{\hat{E}^L_{\bar{a}}}\hat{E}^R_a
     = \Bgen{\hat{E}^L_{\bar{a}}}{\hat{E}^R_a} &= 0 ,
\end{aligned}
\end{equation}
and we see the $\su(2)\times\su(2)$ algebra explicitly.

\subsubsection{Relation to gauged supergravity}

It is known that there is a consistent truncation of type IIA
supergravity on $S^3$~\cite{CLP,CLPST2} giving a maximal $\SO(4)$
gauged supergravity in seven dimensions\footnote{Group manifolds
   always give consistent truncations~\cite{DNP}, but viewing $S^3$ as
   $\SU(2)$ would only give an $\SU(2)$ gauging, whereas here the full
   $\SO(4)$ group is gauged.}.
Making a further consistent truncation to the NSNS fields gives a
half-maximal $\SO(4)$ gauged theory. The embedding tensor of the
half-maximal gauged supergravity~\cite{half-embed,embed-lecture} is a
three-form $X_{ABC}$ where $A=1,\dots 6$ labels an $\SO(3,3)$ vector
index. If one raises one index with the $O(3,3)$ metric one can regard
$X_{AB}{}^C=(X_A)_B{}^C$ as a set of  $\so(3,3)$ matrices labelled by
the index $A$. To define a  gauged supergravity one requires the
quadratic constraint~\cite{embed-lecture}
\begin{equation}
\label{eq:quad-cond}
   \BLie{X_A}{X_B} = - X_{AB}{}^C X_C .
\end{equation}

In terms of the generalised geometry $X$ is encoded in the frame
algebra~\eqref{eq:embed-def}. The quadratic condition simply
follows from the Leibniz property of the generalised Lie
derivative and $X$ can be interpreted as the generalised torsion of
the unique generalised derivative $\hat{D}$ satisfying
$\hat{D}\hat{E}_A=0$~\cite{csw2} (see also
appendix~\ref{app:gen-conn}). This is again in complete analogy with
the conventional geometrical structure of a local group manifold --
there is a unique torsionful connection (the Weitzenb\"ock connection) 
satisfying $\hat{\nabla}\hat{e}_a=0$ such that the torsion of
$\hat{\nabla}$ equals the structure constants of the Lie algebra. As
in the conventional case, the generalised version $\hat{D}$, discussed
in~\cite{gen-weitz1,gen-weitz2}, can be defined if and only if the
space is generalised parallelisable.   

For the $S^3$ parallelisation, we see from~\eqref{eq:su2su2} that
\begin{equation}
   X_{abc} = R^{-1} \epsilon_{abc} , \qquad
   X_{\bar{a}\bar{b}\bar{c}}
      = R^{-1}\epsilon_{\bar{a}\bar{b}\bar{c}} ,
\end{equation}
with all other components vanishing. In $\SL(4,\bbR)$ indices the
self-dual and anti-self dual parts of $X_{ABC}$ correspond to $X_{ij}$
and $X^{\prime ij}$ and we have $X_{ij}=R^{-1}\delta_{ij}$. This
indeed matches the known embedding tensor for the $\SO(4)$
theory~\cite{DLR,DFMR}.

We can also identify the scalar fields of the truncated theory. Given
the frame is always required to be orthonormal with respect to the
$\SO(d,d)$ metric, that is $\eta(\hat{E}'_A,\hat{E}'_B)=\eta_{AB}$,
the scalar fields $U_A{}^B$ in the generalised Scherk--Schwarz
reduction~\eqref{eq:gen-SS} parameterise an
$\SO(d,d)/\SO(d)\times\SO(d)$ coset. Specialising to the $S^3$ case,
and using $\GL^+(4,\bbR)$ indices we can follow the discussion of
section~\ref{sec:gen-SS}. We find the form of the metric and $B$-field
from~\eqref{eq:scalars}
\begin{equation}
\begin{aligned}
   \dd s^{\prime 2} &= \frac{R^2}{T^{kl}y_ky_l}
         \; T^{-1}_{ij} \dd y^i \dd y^j , \\
   B' &= \frac{R^2}{2(T^{kl}y_ky_l)}\; \epsilon_{i_1i_2i_3i_4}
       (T^{i_1j}y_j)y^{i_2} \dd y^{i_3} \wedge \dd y^{i_4} + B , \\
   \ee^{2\Delta'} &= 1 . 
\end{aligned}
\end{equation}
We see that the warp-factor $\Delta'$ is trivial and the metric and
$B$-field scalar dependence on $T$ matches exactly that for the $S^3$
consistent truncation in~\cite{CLP,CLPST2}.

\subsubsection{Other parallelisations}
\label{sec:other-para}

It is interesting to note that other parallelisations of $E$ exist,
and give different gaugings and truncation ansatze on the same
round $S^3$ space. In particular, we could choose a frame based solely
on the left-invariant vectors and one-forms
\begin{equation}
\begin{aligned}
   \hat{E}^L_{\bar{a}}
       &= l_{\bar{a}} - \lambda_{\bar{a}} - i_{l_{\bar{a}}}B , \\
   \hat{E}^R_a &= l_a + \lambda_a - i_{l_a}B .
\end{aligned}
\end{equation}
The algebra now reads
\begin{equation}
\begin{aligned}
  \Lgen_{\hat{E}^L_{\bar{a}}}\hat{E}^L_{\bar{b}}
     = \Bgen{\hat{E}^L_{\bar{a}}}{\hat{E}^L_{\bar{b}}}
     &= R^{-1}\epsilon_{\bar{a}\bar{b}\bar{c}}
        \hat{E}^L_{\bar{c}} , \\
  \Lgen_{\hat{E}^R_a}\hat{E}^R_b
     = \Bgen{\hat{E}^R_a}{\hat{E}^R_b}
     &= R^{-1}\epsilon_{ab\bar{c}} \hat{E}^L_{\bar{c}} , \\
  \Lgen_{\hat{E}^R_a}\hat{E}^L_{\bar{a}}
     = \Bgen {\hat{E}^R_a}{\hat{E}^L_{\bar{b}}}
     &= R^{-1}\epsilon_{a\bar{b}\bar{c}} \hat{E}^L_{\bar{c}} .
\end{aligned}
\end{equation}
This is clearly a different gauging, not isomorphic under $\SO(3,3)$
transformations to the $\SO(3)\times\SO(3)$ gauging of the previous
section, since the embedding tensor $X_{MNP}$ is now not
self-dual. Instead it defines an $\SO(3)$ gauging~\cite{DLR,DFMR}.

This is really a convention flux compactification on a group manifold,
where $l_a$ defines the conventional parallelisation. To match the
usual description, we can fix a different convention for the
generalised frame, taking the linear combinations
\begin{equation}
\label{eq:Odd-split}
   \hat{E}_A = \begin{cases}
      \hat{E}_a = \tfrac{1}{2}\big(\hat{E}^R_a + \hat{E}^L_a\big)
         = l_a - i_{l_a} B , \\
      \hat{\tilde{E}}^a
         = \tfrac{1}{2}\big(\hat{E}^{Ra} - \hat{E}^{La}\big)
         = \lambda^a ,
      \end{cases}
\end{equation}
such that $\eta$ takes the form
\begin{equation}
\label{eq:Odd-norm}
   \eta(\hat{E}_A,\hat{E}_B)
      = \frac{1}{2}\begin{pmatrix}
            0 & \delta_a{}^b \\ \delta^a{}_b & 0
         \end{pmatrix} .
\end{equation}
The algebra then reads
\begin{equation}
\begin{aligned}
   \Bgen{\hat{E}_a}{\hat{E}_b}
     &= f_{ab}{}^c \hat{E}_c + H_{abc}{\tilde{E}}^c , \\
   \Bgen{\hat{E}_a}{\hat{\tilde{E}}^b}
     &= - f_{ac}{}^b \hat{\tilde{E}}^c , \\
   \Bgen{\hat{\tilde{E}}^a}{\hat{\tilde{E}}^b} &= 0 ,
\end{aligned}
\end{equation}
where
\begin{equation}
   f_{ab}{}^c = R^{-1}\epsilon_{ab}{}^c , \qquad
   H_{abc} = R^{-1} \epsilon_{abc} ,
\end{equation}
As usual, $f_{ab}{}^c$ characterises the Lie algebra of the group
manifold (here $\su(2)$) and $H=\frac{1}{6}H_{abc}l^a\wedge l^b\wedge
l^c$ is the three-form flux~\cite{KM,HR-E1,GMPW}.


\subsection{$S^4$ and $E_{4(4)}$ generalised geometry}
\label{sec:s4}

We next consider the $\AdS_7\times S^4$ solution~\cite{FR,PvNT} of
eleven-dimensional supergravity
\begin{equation}
\begin{aligned}
   \dd s^2 &= \dd s^2(\AdS_7) + R^2\dd s^2(S^4) , \\
   F &= 3R^{-1} \vol_g , 
\end{aligned}
\end{equation}
where $R$ is the radius of the four-sphere and we are using the
conventions of~\cite{csw2,csw3}. That this theory has a consistent
truncation to seven dimensions has been proven by Nastase, Vaman and
van~Nieuwenhuizen~\cite{NVvN}.

In terms of the $\GL^+(5,\bbR)$ generalised geometry on the $S^4$ we
have
\begin{equation}
   E \simeq TM \oplus \Lambda^2T^*M .
\end{equation}
However this is precisely the generalised (exceptional)
geometry in four dimensions~\cite{chris}, where we identify the
U-duality exceptional group and its maximally compact subgroup
\begin{equation}
   E_{4(4)}\times \bbR^+ \simeq \GL^+(5,\bbR)
   \quad \text{and} \quad
   H_4 \simeq \SO(5) .
\end{equation}
This geometry was discussed in the context of an extension of Double
Field Theory in~\cite{BP-4d,BP-alg} and in the general context of
exceptional generalised geometry and generalised curvatures
in~\cite{csw2,csw3}.

The embedding tensor $X_{AB}{}^C$ in this case transforms in the
$\rep{15}+\rep{40}$ representation of $\SL(5,\bbR)$~\cite{SW}. From
the form of the frame algebra~\eqref{eq:SO-alg}, one finds that the
two components are given by
\begin{equation}
   X_{ij} = R^{-1}\delta_{ij} , \qquad
   X_{ijk}{}^l = 0 ,
\end{equation}
which reproduces the standard embedding tensor of maximal
seven-dimensional $\SO(5)$ gauged supergravity~\cite{PPvN}. The scalar
field ansatz is given by~\eqref{eq:scalars} where $A$ is a
three-form. Again this agrees with the ansatz derived in~\cite{NVvN}.


\subsection{$S^5$ and $E_{6(6)}$ generalised geometry}
\label{sec:s5}

We next consider the $\AdS_5\times S^5$ solution~\cite{Schwarz} of Type
IIB supergravity
\begin{equation}
\label{eq:S5-sol}
\begin{aligned}
   \dd s^2 &= \dd s^2(\AdS_5) + R^2\dd s^2(S^5) , \\
   F &= 4R^{-1} \left( \vol_g + \vol_{\AdS} \right) ,
\end{aligned}
\end{equation}
where $R$ is the radius of the five-sphere, $\vol_g$ is the volume
form on $S^5$, $\vol_{\AdS}$ is the volume form on $\AdS_5$ and $F$ is
the self-dual five-form RR flux. We are using the conventions
of~\cite{IIB-conv} for the type IIB supergravity.

If we keep the full degrees of freedom of the Type IIB theory, the
$\GL^+(6,\bbR)$ generalised geometry embeds in a larger
(exceptional) $E_{6(6)}\times\bbR^+$ generalised
geometry~\cite{chris,csw2}. This is summarised in 
appendix~\ref{app:e6-gen-geom}, partly using results of
Ashmore~\cite{anthony}. One considers the 27-dimensional generalised
tangent space~\cite{chris}
\begin{equation}
\label{eq:E6E}
\begin{aligned}
   E &\simeq TM \oplus (T^*M \oplus T^*M) \oplus \Lambda^3T^*M
      \oplus (\Lambda^5T^*M \oplus \Lambda^5T^*M) , \\
   V &= v + \rho^\xa + \lambda + \chi^\xa .
\end{aligned}
\end{equation}
where $\xa$ labels a doublet of the IIB S-duality $\SL(2,\bbR)$
group. There is an natural action of $E_{6(6)}\times\bbR^+$ on $V\in
E$ that preserves the symmetric top-form cubic invariant~\cite{anthony}
\begin{equation}
   c(V,V,V) = \tfrac{1}{2}i_v\lambda \wedge \lambda
       + \tfrac{1}{2} \lambda\wedge\rho_\xa\wedge\rho^\xa
       +  (i_v\rho_\xa) \chi^\xa \in \Lambda^6T^*M ,
\end{equation}
where we lower $\SL(2,\bbR)$ indices by
$u_\xa=\epsilon_{\xa\xb}u^\xb$. For $V,V'\in E$ there is a generalised
Lie derivative~\cite{csw2,anthony}, just as
in~\eqref{eq:Lgen} but now such that $\oadj$ projects onto the
$E_{6(6)}\times\bbR^+$ adjoint representation,
\begin{equation}
\label{eq:Lgen-e6-*}
\begin{aligned}
   \Lgen_V V' &= (V\cdot\der)V' - (\der \oadj V)V' \\
      &= \BLie{v}{v'}
      + \mathcal{L}_v\rho^{\prime \xa}
      - i_{v'}\dd \rho^\xa
      + \mathcal{L}_v\lambda - i_{v'}\dd\lambda
          + \dd\rho_\xa\wedge\rho^{\prime \xa} \\ & \qquad
      + \mathcal{L}_v\chi^{\prime\xa}
          - \dd\lambda\wedge\rho^{\prime\xa}
          + \dd\rho^{\xa}\wedge\lambda' .
\end{aligned}
\end{equation}
This captures diffeomorphisms together with the type IIB gauge
transformations of NSNS and RR fields.

There is also a generalised metric $G$ which is invariant under the
maximal compact subgroup $H_6=\USp(8)/\bbZ_2\subset E_{6(6)}\times\bbR^+$
and unifies all the bosonic degrees of freedom along with the warp
factor $\Delta$ of the non-compactified space. The corresponding
generalised orthonormal frame $\{\hat{E}_A\}$ transforms in the
$\rep{27}$ representation of $\USp(8)$. For what follows we can
actually use the decomposition under the subgroup
$\SO(6)\times\SO(2)\simeq \SU(4)/\bbZ_2\times\SO(2)
\subset\USp(8)/\bbZ_2$, giving
\begin{equation}
\begin{aligned}
   \{\hat{E}_A\}
       &= \{\hat{E}_{ij}\} \cup \{\hat{E}^i_\za\} ,\\
   \rep{27} &= (\rep{15},\rep{1}) + (\rep{6},\rep{2}) ,
\end{aligned}
\end{equation}
where $i=1,\dots,6$ and $\za=1,2$. The orthonormal condition reads
\begin{equation}
\label{eq:E6-ortho}
\begin{aligned}
   G(\hat{E}_{ij},\hat{E}_{kl})
      &= \delta_{ik}\delta_{jl} - \delta_{il}\delta_{jk} , \\
   G(\hat{E}_{ij},\hat{E}^{ak})
      &= 0 , \\
   G(\hat{E}^i_\za,\hat{E}^j_\zb)
      &= \delta_{\za\zb}\delta^{ij} .
\end{aligned}
\end{equation}
Given the isomorphism~\eqref{eq:E6E} we can again define a sub-class
of orthonormal frames that transform under an $\SO(5)$ subgroup of
$\SO(6)\subset\USp(8)/\bbZ_2$. The corresponding ``split'' generalised
frame $\{\hat{E}_A\}$, analogous to~\eqref{eq:split}, can be written
as
\begin{equation}
\label{eq:E6-split-1}
\begin{aligned}
   \hat{E}_{ij}
      &= \begin{cases}
         \hat{E}_{a6} = \hat{E}_a , \\
         \hat{E}_{ab}
           = \frac{1}{3!}\epsilon_{abc_1c_2c_3}\hat{E}^{c_1c_2c_3},
         \end{cases} , & &&
   \hat{E}_\za^i
      &= \begin{cases}
         \hat{E}_\za^a = \hat{E}_\za^a , \\
         \hat{E}_\za^6 = \hat{E}^{12345}_\za
         \end{cases} ,
\end{aligned}
\end{equation}
where
\begin{equation}
\label{eq:E6-split-2}
\begin{aligned}
   \hat{E}_a
      &= \ee^\Delta\big(
         \hat{e}_a - i_{\hat{e}_a} B^\xa - i_{\hat{e}_a} A
         - \tfrac{1}{2}B_\xa\wedge i_{\hat{e}_a} B^\xa
         \\ & \qquad \qquad \qquad
         - B^\xa \wedge i_{\hat{e}_a} A
         - \tfrac{1}{6}B^\xa\wedge B_\xb\wedge i_{\hat{e}_a}B^\xb
         \big), \\
   \hat{E}^a_\za
      &= \ee^\Delta\ee^{-\phi/2}\big(
         \hat{f}_\za{}^\xa e^a + B_\za\wedge e^a
         - \hat{f}_\za{}^\xa A \wedge e^a
         + \tfrac{1}{2}B^\xa \wedge B_\za\wedge e^a \big) ,
         \\
   \hat{E}^{abc}
      &= \ee^\Delta\ee^{-\phi} \left(
         e^{abc} + B^\xa\wedge e^{abc}\right) ,
         \\
   \hat{E}^{a_1\dots a_5}_\za
      &= \ee^\Delta\ee^{-3\phi/2} \hat{f}_\za{}^\xa e^{a_1\dots a_5} .
\end{aligned}
\end{equation}
We have the usual $\SL(2,\bbR)$ frame
\begin{equation}
\label{eq:f-def}
   \hat{f}_\za{}^\xa
       = \begin{pmatrix}
             \ee^{\phi/2} & C \ee^{\phi/2} \\  0 & \ee^{-\phi/2}
          \end{pmatrix} ,
\end{equation}
and define $B_\za=\hat{f}_\za{}^\xa
B_\xa=\hat{f}_\za{}^\xa\epsilon_{\xa\xb}B^\xb$ and $e^{a_1\dots a_n}
=e^{a_1}\wedge\dots\wedge e^{a_n}$. The split frame encodes the
string-frame metric $g$, dilaton $\phi$ and warp factor $\Delta$,
while the NSNS two-form is given by $B^1$ and the RR form field
potentials are $C_{(0)}=C$, $C_{(2)}=B^2$, and $C_{(4)}=A$. Note that
the inverse generalised metric can be written as
\begin{equation}
   G^{-1\, MN} = \delta^{AB}\hat{E}_A^M\hat{E}_B^N
      = \tfrac{1}{2}\delta^{ik}\delta^{jl}
          \hat{E}_{ij}^M \hat{E}_{jk}^N
        + \delta^{\za\zb}\delta_{ij}
           \hat{E}_\za^{i\,M} \hat{E}_\zb^{j\,N} .
\end{equation}
Certain components of $G^{-1}$ are given explicitly
in~\eqref{eq:G-comp}.

For the application to $S^5$ we are interested in structures
defined by the subgroups
\begin{equation}
\begin{aligned}
   E_{6(6)}\times\bbR^+ & \supset \GL^+(6,\bbR) \times \SL(2,\bbR), \\
   H_6 = \USp(8)/\bbZ_2 &\supset \SU(4)/\bbZ_2 \times \SO(2)
       \simeq \SO(6)\times\SO(2) ,
\end{aligned}
\end{equation}
where again $\SL(2,\bbR)$ is the S-duality group. We find that the
generalised tangent space decomposes as
\begin{equation}
\begin{aligned}
   E &\simeq E^{(0)} \oplus E^{(\xa)} , \\
   \rep{27} &= \repp{15}{1} + \repp{6'}{2} ,
\end{aligned}
\end{equation}
where
\begin{equation}
   E^{(0)} \simeq TM \oplus \Lambda^3T^*M , \qquad
   E^{(\xa)} \simeq T^*M \oplus \Lambda^5T^*M .
\end{equation}
Comparing with~\eqref{eq:fund-rep} we see that $E^{(\xa)}\simeq (\det
T^*M)^{1/2}\otimes W^*$. This means that it is a
$\GL^+(6,\bbR)\simeq\bbR^+\times\SL(6,\bbR)$ one-form weighted by a
$\bbR^+$ factor of $(\det T^*M)^{-1/2}$.

We now show that the $S^5$ solution actually gives a parallelisation
of the full tangent space $E$. We define a frame
\begin{equation}
\label{eq:E6para}
   \hat{E}_A = \begin{cases}
      \hat{E}_{ij} = v_{ij} + \sigma_{ij} - i_{v_{ij}} A
         & \quad \text{for } E^{(0)}, \\
      \hat{E}^i_\za
         = \hat{f}_\za{}^\xa \left( R\dd y^i + y^i \vol_g + R\dd y^i
            \wedge A \right)
         & \quad \text{for $E^{(\xa)}$} ,
      \end{cases}
\end{equation}
where the $\SL(2,\bbR)$ frame is simply~\eqref{eq:f-def} with
\emph{constant} dilaton $\phi$ and RR scalar $C$. Since the $E^{(0)}$
component is exactly of the type discussed in section~\ref{sec:S-GG}
we just use the frame~\eqref{eq:globalE} which we know is globally
defined. For $E^{(a)}$ we note that $\dd y_i$ vanishes on $y_i^2=1$ so
the frame is nowhere vanishing (and is essentially the dual of the
$\hat{E}_i$ frame on $W$). It is easy to see that the parallelising
frame~\eqref{eq:E6para} is orthonormal,
satisfying~\eqref{eq:E6-ortho}, for the round sphere with flux
background~\eqref{eq:S5-sol}.

We can again work out the algebra of the frame under the generalised
Lie derivative~\eqref{eq:Lgen-e6-*}. Since $\hat{E}^{ai}$ is closed this
reduces to using the generalised Lie derivative for the
$\GL^+(6,\bbR)$ subgroup. We find
\begin{equation}
\label{eq:E6-SO6}
\begin{aligned}
   L_{\hat{E}_{ij}} \hat{E}_{kl}
      &= R^{-1} \big(\delta_{ik}\hat{E}_{jl}
          - \delta_{il}\hat{E}_{jk}
          - \delta_{jk}\hat{E}_{il}
          + \delta_{jl}\hat{E}_{ik} \big) , \\
   L_{\hat{E}_{ij}} \hat{E}_\za^k
      &= R^{-1} \big(\delta_{il}\delta_j^k\hat{E}_\za^l
          - \delta_{jl}\delta_i^k\hat{E}_\za^l \big) , \\
   L_{\hat{E}_\za^i} \hat{E}_{jk} &= 0 , \\
   L_{\hat{E}_\za^i} \hat{E}_\zb^j &= 0 .
\end{aligned}
\end{equation}
Note that unlike the previous examples we have
\begin{equation}
   \Lgen_{\hat{E}_A} \hat{E}_B \neq \Bgen{\hat{E}_A}{\hat{E}_B} ,
\end{equation}
and~\eqref{eq:E6-SO6} does not define a Lie algebra but rather a
Leibniz algebra.

\subsubsection{Consistent truncations and the general scalar ansatz on
   $S^5$}
\label{sec:e6-truncation}

It is widely believed that there is a consistent truncation on $S^5$
to an $\SO(6)$ maximally supersymmetric $d=5$ supergravity. The metric
and five-form flux subsector was shown to be consistent
in~\cite{CLPST1,CLP}, but otherwise there is no complete derivation of
consistency. In the following we will show that generalised
parallelisable structure~\eqref{eq:E6para} reproduces the correct
gauge structure and matches the known scalar ansatz for $g_{mn}$ and
$A_{m_1\dots m_4}$. Furthermore we will derive the full scalar ansatz
including the remaining bosonic fields.

The embedding tensor $T_{AB}{}^C$ of five-dimensional
maximally supersymmetric supergravity transforms in the $\rep{351}$
representation of $E_{6(6)}$~\cite{5D-gauge}. Decomposing under
$\SL(6,\bbR)\times\SL(2,\bbR)$ this splits as
\begin{equation}
   \rep{351} = (\rep{21},\rep{1}) + (\rep{15},\rep{3})
      + (\bar{\rep{84}},\rep{2}) + (\bar{\rep{6}},\rep{2})
      + (\rep{105},\rep{1}) .
\end{equation}
For the $\SO(6)$ gauging, only the $(\rep{21},\rep{1})$ component is
non-zero. Specifically, decomposing the $E_{6(6)}$ index as
$A=\{ii',\za i\}$, one has
\begin{equation}
\begin{aligned}
   X_{ii',jj'}{}^{kk'}
      &= X_{ij}\delta^{kk'}_{i'j'}
         - X_{i'j}\delta^{kk'}_{ij'}
         - X_{ij'}\delta^{kk'}_{i'j}
         + X_{i'j'}\delta^{kk'}_{ij} , \\
   X_{ii',\zb,k}^{\ph{ii',}j,\zc}
      &=  \left( X_{ik}\delta^j_{i'}
         - X_{i'k}\delta^j_{i}\right) \delta^\zc_\zb ,
\end{aligned}
\end{equation}
with all other components vanishing. We see that the
algebra~\eqref{eq:E6-SO6} corresponds to $X_{ij}=R^{-1}\delta_{ij}$ in
agreement with the standard $\SO(6)$ gauging embedding
tensor~\cite{5D-gauge}.

The scalar fields in the truncation enter via the usual
Scherk--Schwarz rotation
\begin{equation}
   \hat{E}'_A(x) = U_A{}^B(x) \hat{E}_B , \qquad \qquad
   U = \begin{pmatrix}
         U_{ii'}{}^{jj'} & U_{ii',j}^{\ph{ii',}\zb} \\
         U_\za^{i,jj'} & U_{\za,j}^{i,\zb}
      \end{pmatrix} \in E_{6(6)} .
\end{equation}
Note that under $\GL^+(6,\bbR)\times\SL(2,\bbR)$, given a generalised
vector $V^A=(V^{ii'},V^\za_i)$ the cubic invariant is given
by~\cite{anthony}
\begin{equation}
   c(V,V,V)
      = \tfrac12 \tfrac{1}{6!} \epsilon_{i_1\dots i_6} V^{i_1i_2}V^{i_3i_4}V^{i_5i_6}
         + \tfrac12 V^{ij} V_{\za i} V^\za_j .
\end{equation}
and $U$ is defined as the transformation that leaves $c$
invariant. Unlike the previous cases, we cannot easily parameterise the
coset $E_{6(6)}/\USp(8)$. However, comparing~\eqref{eq:G-comp} for the
split frame~\eqref{eq:E6-split-1} we can read off expressions for the
metric and potentials
\begin{equation}
\begin{aligned}
   \ee^{2\Delta'} g^{\prime mn}
      &= \delta^{AB} U_A{}^{jj'} U_B{}^{kk'} v_{jj'}^m v_{kk'}^n , \\
   \ee^{2\Delta'} B^{\prime\xa}_{mn}
      &= \delta^{AB} U_A{}^{jj'} U_{B\,k}^{\ph{B\,}\zc} \hat{f}_\zc{}^\xa
          R \;v_{jj'\,[m}  \der_{n]} y^k , \\
   \ee^{2\Delta'} \left( A'_{mnpq}
         - \tfrac{3}{2}B'_{\xa m[n} B^{\prime\xa}_{pq]} - A_{mnpq}
         \right)
      &= - \delta^{AB} U_A{}^{jj'} U_B{}^{kk'}
          v_{jj'\,m} \lambda_{kk'\,npq} ,
\end{aligned}
\end{equation}
where the spacetime index on $v_{jj'}$ in the last two expressions is
lowered using $g'_{mn}$. Note that totally antisymmetrizing the final
expression eliminates the $B^{\prime\xa}$ terms. Comparing
with~\eqref{eq:G-comp2} we also find
\begin{equation}
   \ee^{2\Delta'}\big(
         \ee^{-\phi'} h^{\prime\xa\xb}g'_{mn}
             - B^{\prime\xa}_{mp}g^{\prime pq}B^{\prime\xb}_{qn}
         \big)
      = \delta^{AB} U_{A\,i}^{\ph{A\,}\za} U_{B\,j}^{\ph{B\,}\zb}
           \hat{f}_\za{}^\xa \hat{f}_\zb{}^\xb
           R^2 \der_my^i \der_ny^j ,
\end{equation}
which defines the new $\SL(2,\bbR)$ metric $h^{\prime\xa\xb}$. These
expressions are the direct analogues of those derived for the $S^7$
truncation in~\cite{dWN-S7,dWN,N-new}.

If one specialises to the case where $U_A{}^B$ parameterises only the
$\SL(6,\bbR)/\SO(6)$ subspace, that is take
$U_{ii'}{}^{ja}=U_{ia}{}^{jj'}=0$ and
$U_{ai}{}^{bk}=\delta_a^b\delta_i^j$, we are back to the case
discussed in section~\ref{sec:S-GG}. The two-form fields vanish,
$B^\alpha=0$, $\phi$ and $C$ are already moduli, and in addition 
\begin{equation}
\begin{aligned}
   \dd s^{\prime 2} &= \frac{R^2}{(T^{kl}y_ky_l)^{1/2}}
         \; T^{-1}_{ij} \dd y^i \dd y^j , \\
   A' &= \frac{1}{2(T^{kl}y_ky_l)} \frac{R^4}{3!} \;\epsilon_{i_1\dots i_6}
       (T^{i_1j}y_j)y^{i_2} \dd y^{i_3}\wedge
          \dots \wedge \dd y^{i_6} + A , \\
   \ee^{2\Delta'} &= (T^{kl}y_ky_l)^{1/2}
\end{aligned}
\end{equation}
which is in complete agreement with the ansatz of~\cite{CLPST1,CLP}.


\subsection{$S^7$ and $E_{7(7)}$ generalised geometry}
\label{sec:s7}

We next consider the $\AdS_7\times S^4$ solution~\cite{FR} of
eleven-dimensional supergravity
\begin{equation}
\begin{aligned}
   \dd s^2 &= \dd s^2(\AdS_4) + R^2\dd s^2(S^7) , \\
   \tilde{F} &= 6R^{-1} \vol_g ,
\end{aligned}
\end{equation}
where $R$ is the radius of the seven-sphere and $\tilde{F}$ is the
seven-form flux, that is the eleven-dimensional dual of the usual
four-form. Here we are using the conventions of~\cite{csw2,csw3}. It
is a classic result due to de~Wit and Nicolai~\cite{dWN-S7,NP} that
this background admits a consistent truncation to $\SO(8)$ gauged
$N=8$ supergravity in four dimensions.

Here we will give a new interpretation of this truncation in terms of
generalised geometry. The generalised frame~\eqref{eq:E7-split-*}
described below has, in fact, already appeared in the work
of~\cite{N-new} as has the form of the scalar
ansatz~\cite{dWN-S7,dWNW,N-new}. However, the key new points are,
first, to note that this frame is a parallelisation of the generalised
tangent space and, second, that the $\SO(8)$ embedding tensor is
encoded in the frame algebra under the generalised Lie
derivative. This shows that the truncation actually falls within the
class of generalised Scherk--Schwarz reductions. 

If we keep all the degrees of freedom of the eleven-dimensional
supergravity, we are led to an $\E7\times\bbR^+$ (exceptional)
generalised geometry. One considers the generalised tangent
space~\cite{chris,PW}
\begin{equation}
\label{eq:E7E}
\begin{aligned}
   E &\simeq TM \oplus \Lambda^2T^*M \oplus \Lambda^5T^*M
         \oplus (T^*M \otimes\Lambda^7T^*M) , \\
   V &= v + \omega + \sigma + \tau ,
\end{aligned}
\end{equation}
which transforms as the $\rep{56}_{\rep{1}}$ representation under
$E_{7(7)}\times\bbR^+$ action, where a scalar $\rep{1}_{\rep{k}}$ of
weight $k$ under $\bbR^+$ is a section of $(\det T^*M)^{k/2}$. Given
$V,V'\in E$ there is a generalised Lie derivative~\cite{csw2}
given by\footnote{The ``$j$-notation'' for the $T^*M
   \otimes\Lambda^7T^*M$ component is described in~\cite{PW,csw2,csw3}.}
\begin{equation}
\label{eq:Lgen-e7}
\begin{aligned}
   \Lgen_V V' &= (V\cdot\der)V' - (\der \oadj V)V' \\
      &= \mathcal{L}_v v'
          + \left( \mathcal{L}_v \omega' - i_{v'} \dd\omega \right)
          + \left( \mathcal{L}_v \sigma' - i_{v'} \dd\sigma
          - \omega'\wedge\dd\omega \right)
       \\ & \qquad
          + \left( \mathcal{L}_v \tau'
             - j\sigma'\wedge\dd\omega
             - j\omega'\wedge\dd\sigma \right) ,
\end{aligned}
\end{equation}
which captures diffeomorphisms together with the gauge
transformations of three-form and dual six-form gauge fields. There
is also a generalised metric which is invariant under the maximal
compact subgroup $H_7=\SU(8)/\bbZ_2$ and unifies all the bosonic degrees of
freedom along with the warp factor $\Delta$.  The corresponding
generalised orthonormal frame $\{\hat{E}_A\}$ transforms in the
complex two-form $\rep{28}_\bbC$ representation of $\SU(8)$. For what
follows we can actually use the decomposition under the subgroup
$\SO(8)\subset\SU(8)/\bbZ_2$, giving
\begin{equation}
\begin{aligned}
   \{\hat{E}_A\}
       &= \{\hat{E}_{ij}\} \cup \{\hat{E}^{\prime ij} \} ,\\
   \rep{28}_\bbC &= \rep{28} + \rep{28} .
\end{aligned}
\end{equation}
The orthonormal condition reads
\begin{equation}
\begin{aligned}
   G(\hat{E}_{ij},\hat{E}_{kl})
      &= \delta_{ik}\delta_{jl} - \delta_{il}\delta_{jk} , \\
   G(\hat{E}_{ij},\hat{E}^{\prime\, kl})
      &= 0 , \\
   G(\hat{E}^{\prime\, ij},\hat{E}^{\prime\, kl})
      &= \delta^{ik}\delta^{jl} - \delta^{il}\delta^{jk}
\end{aligned}
\end{equation}
Note that the full $\SU(8)$ representation and its conjugate have the
form
\begin{equation}
\label{eq:SU8-basis}
\begin{aligned}
   \hat{E}_{\alpha\beta}
      &= -\tfrac{1}{32}\ii \gamma^{ij}_{\alpha\beta}
         \big( \hat{E}_{ij} - \ii \hat{E}'_{ij} \big) , \\
   \bar{\hat{E}}^{\alpha\beta}
      &= \tfrac{1}{32}\ii \gamma^{ij\,\alpha\beta}
         \big( \hat{E}_{ij} + \ii \hat{E}'_{ij} \big) ,
\end{aligned}
\end{equation}
where $\gamma^{ij}_{\alpha\beta}$ are $\Spin(8)$ gamma matrices,
$\alpha,\beta=1,\dots,8$ are $\SU(8)$ indices and we are matching the
conventions of~\cite{csw2,csw3}. Given the isomorphism~\eqref{eq:E7E}
we can again define a sub-class of orthonormal frames that transform
under an $\SO(7)$ subgroup of $\SO(8)\subset\SU(8)/\bbZ_2$. The
corresponding ``split'' generalised frame $\{\hat{E}_A\}$, analogous
to~\eqref{eq:split}, can be written as
\begin{equation}
\label{eq:E7-split-*}
\begin{aligned}
   \hat{E}_{ij}
      &= \begin{cases}
         \hat{E}_{a7} = \hat{E}_a , \\
         \hat{E}_{ab}
           = \frac{1}{5!}\epsilon_{abc_1\dots c_5}
               \hat{E}^{c_1\dots c_5} ,
         \end{cases} , & &&
   \hat{E}^{\prime ij}
      &= \begin{cases}
         \hat{E}^{\prime a7} = \hat{E}^{a,12\dots 7} , \\
         \hat{E}^{\prime ab} = \hat{E}^{ab} ,
         \end{cases} ,
\end{aligned}
\end{equation}
where
\begin{equation}
\label{eq:E7-split}
\begin{aligned}
   \hat{E}_a &= \ee^{\Delta} \Big( \hat{e}_a + i_{\hat{e}_a} A
      + i_{\hat{e}_a}\tA
      + \tfrac{1}{2}A\wedge i_{\hat{e}_a}A
      \\ & \qquad \qquad
      + jA\wedge i_{\hat{e}_a}\tA
      + \tfrac{1}{6}jA\wedge A \wedge i_{\hat{e}_a}A \Big) , \\
   \hat{E}^{ab} &= \ee^\Delta \left( e^{ab} + A\wedge e^{ab}
      - j\tA\wedge e^{ab}
      + \tfrac{1}{2}jA\wedge A \wedge e^{ab} \right) , \\
   \hat{E}^{a_1\dots a_5} &= \ee^{\Delta} \left( e^{a_1\dots a_5}
      + jA\wedge e^{a_1\dots a_5} \right) , \\
   \hat{E}^{a,a_1\dots a_7} &= \ee^\Delta e^a\otimes e^{a_1\dots a_7} ,
\end{aligned}
\end{equation}
where $e^{ab}=e^a\wedge e^b$ etc. and $A$ and $\tilde{A}$ are the
three- and dual six-form potentials respectively. This particular form
of $E_{7(7)}$ frame first appeared in an extended $(4+56)$-dimensional
formulation of eleven-dimensional supergravity in~\cite{hillmann}. It
arose via a non-linear realisation of $\E7$, following an embedding in
$E_{11}$, in~\cite{BGPW} and in generalised geometry
in~\cite{csw2,csw3}. It recently appeared in the context of extending
the original de~Wit--Nicolai analysis of~\cite{dWN}
in~\cite{N-new}.

For the current application to $S^7$ we are interested in structures
defined by the subgroups
\begin{equation}
\begin{aligned}
   E_{7(7)}\times\bbR^+ & \supset \GL^+(8,\bbR) , \\
   H_7 \simeq \SU(8)/\bbZ_2 &\supset \SO(8)/\bbZ_2 .
\end{aligned}
\end{equation}
We find that the generalised tangent space decomposes under
$\GL^+(8,\bbR)$ as
\begin{equation}
\begin{aligned}
   E &\simeq E^{(0)} \oplus E^{(1)} , \\
   \rep{56} &= \rep{28} + \rep{28}' , 
\end{aligned}
\end{equation}
where
\begin{equation}
   E^{(0)} \simeq TM \oplus \Lambda^5T^*M , \qquad
   E^{(1)} \simeq \Lambda^2T^*M
       \oplus \left( T^*M\otimes\Lambda^7T^*M \right) .
\end{equation}

We now show that the $S^7$ solution actually gives a parallelisation
of the full tangent space $E$. We define a frame
\begin{equation}
\label{eq:E7para}
   \hat{E}_A = \begin{cases}
      \hat{E}_{ij} = v_{ij} + \sigma_{ij} + i_{v_{ij}} \tilde{A}
         & \quad \text{for } E^{(0)}, \\
      \hat{E}^{\prime\, ij} = \omega_{ij} + \tau_{ij}
           - j\tilde{A}\wedge \omega_{ij}
         & \quad \text{for $E^{(1)}$} ,
      \end{cases}
\end{equation}
where $\omega_{ij}$ and $\tau_{ij}$ are defined
in~\eqref{eq:ost-def}. Note that $\omega_{ij}=0$ when $y_i^2+y_j^2=1$
whereas $\tau_{ij}=0$ when $y_i=y_j=0$ so each $\hat{E}^{\prime ij}$ is
non-vanishing. Furthermore, using the form of the generalised
metric~\cite{csw2,csw3} and~\eqref{eq:contractions} we see that the
frame is orthonormal. Note that the $\SU(8)$ form~\eqref{eq:SU8-basis}
of this frame has already appeared in~\cite{N-new}. 

We can again work out the frame algebra under the generalised Lie
derivative~\eqref{eq:Lgen-e7}. Since $\omega_{ij}$ is closed this
reduces to using the generalised Lie derivative for the
$\GL^+(8,\bbR)$ subgroup. We find 
\begin{equation}
\label{eq:E7-SO8}
\begin{aligned}
   L_{\hat{E}_{ij}} \hat{E}_{kl}
      &= R^{-1} \big(\delta_{ik}\hat{E}_{lj}
          - \delta_{il}\hat{E}_{kj}
          - \delta_{jk}\hat{E}_{li}
          + \delta_{jl}\hat{E}_{ki} \big) , \\
   L_{\hat{E}_{ij}} \hat{E}^{\prime\, kl}
      &= R^{-1} \big(\delta_i^k\delta_{jp} \hat{E}^{\prime\, lp}
          - \delta_i^l\delta_{jp} \hat{E}^{\prime\, kp}
          - \delta_j^k\delta_{ip} \hat{E}^{\prime\, lp}
          + \delta_j^l\delta_{ip} \hat{E}^{\prime\, kp} \big) , \\
   L_{\hat{E}^{\prime\,ij}} \hat{E}_{kl} &= 0 , \\
   L_{\hat{E}^{\prime\,ij}} \hat{E}^{\prime\,kl} &= 0 .
\end{aligned}
\end{equation}
Again, unlike the $S^3$ and $S^4$ examples, we have
\begin{equation}
   \Lgen_{\hat{E}_A} \hat{E}_B \neq \Bgen{\hat{E}_A}{\hat{E}_B} ,
\end{equation}
and~\eqref{eq:E7-SO8} defines a Leibniz algebra. To make the local
$\SU(8)/\bbZ_2$ symmetry more manifest, and hence match more closely
the de~Wit--Nicolai formulation~\cite{dWN-S7,dWN,N-new}, we can use the
combinations~\eqref{eq:SU8-basis}. The frame algebra then reads
\begin{equation}
\begin{aligned}
   \Lgen_{\hat{E}_{\alpha\beta}} \hat{E}_{\gamma\delta}
      &= -\tfrac{1}{8}\ii R^{-1} \Big(
          \delta_{\alpha\gamma}\hat{E}_{\delta\beta}
          - \delta_{\alpha\delta} \hat{E}_{\gamma\beta}
          - \delta_{\beta\gamma} \hat{E}_{\delta\alpha}
          + \delta_{\beta\delta} \hat{E}_{\gamma\alpha}
          \Big) , \\
   \Lgen_{\hat{E}_{\alpha\beta}} \bar{\hat{E}}^{\gamma\delta}
      &= -\tfrac{1}{8}\ii R^{-1} \Big(
      	\delta_\alpha^\gamma
      	\delta_{\beta \epsilon} \bar{\hat{E}}^{ \delta \epsilon}
          - \delta_\beta^\gamma
      	\delta_{\alpha \epsilon} \bar{\hat{E}}^{ \delta \epsilon}
          - \delta_\alpha^\delta
      	\delta_{\beta \epsilon} \bar{\hat{E}}^{\gamma \epsilon}
          + \delta_\beta^\delta
      	\delta_{\alpha \epsilon} \bar{\hat{E}}^{\gamma \epsilon} \Big) . \\
\end{aligned}
\end{equation}

Let us now connect to the consistent truncation. The embedding tensor
$T_{AB}{}^C$ of four-dimensional $N=8$ supergravity
transforms in the $\rep{912}$ representation of
$E_{7(7)}$~\cite{4D-gauge}. Decomposing under $\SL(8,\bbR)$ this
splits as
\begin{equation}
   \rep{912} = \rep{36} + \rep{36}' + \rep{420} + \rep{420}' .
\end{equation}
For the $\SO(8)$ gauging, only the $\rep{36}$ component is
non-zero. Specifically, decomposing the $E_{7(7)}$ index as
as two pairs of indices $ii'$ as in~\eqref{eq:E7para}, we have
\begin{equation}
\begin{aligned}
   X_{ii'\,jj'}{}^{kk'} = - X_{ii'}{}^{kk'}{}_{jj'}
      &= X_{ij}\delta_{i'j'}^{kk'}
         - X_{i'j}\delta_{ij'}^{kk'}
         - X_{ij'}\delta_{i'j}^{kk'}
         + X_{i'j'}\delta_{ij}^{kk'}
\end{aligned}
\end{equation}
with all other components vanishing. We see that the
algebra~\eqref{eq:E7-SO8} corresponds to $X_{ij}=R^{-1}\delta_{ij}$ in
agreement with the standard $\SO(8)$ gauging embedding
tensor~\cite{4D-gauge}.

The scalar fields in the truncation enter via the usual
Scherk--Schwarz rotation. In this case, this was already described
in~\cite{dWN-S7,dWN,N-new}. For completeness, let us include them
here. In our notation, one has
\begin{equation}
   \hat{E}'_A(x) = U_A{}^B(x) \hat{E}_B , \qquad \qquad
   U = \begin{pmatrix}
         U_{ii'}{}^{jj'} & U_{ii'\,jj'} \\
         U^{ii'\,jj''} & U^{ii'}{}_{jj'}
      \end{pmatrix} \in E_{7(7)}
\end{equation}
Following~\cite{N-new}, comparing with the split
frame~\eqref{eq:E7-split} we can read off expressions for the metric
and potentials
\begin{equation}
\begin{aligned}
   \ee^{2\Delta'} g^{\prime mn}
      &= \delta^{AB} U_A{}^{ii'} U_B{}^{jj'} v_{ii'}^m v_{jj'}^n , \\
   \ee^{2\Delta'} A'_{mnp}
      &= \delta^{AB} U_A{}^{ii'} U_{B\,jj'} v_{ii'\,[m} \omega^{jj'}_{np]} , \\
   \ee^{2\Delta'}\Big( \tA'_{m_1 \dots m_6}- \tA_{m_1 \dots m_6} 
         \hspace*{1.7cm} \\
         + \tfrac{1}{2} \tfrac{5!}{2!3!} A'_{m_1 [m_2 m_3} A'_{m_4 m_5 m_6]}
         \Big) 
      &= \delta^{AB} U_A{}^{ii'} U_B{}^{jj'} v_{ii'\,m_1}
         \sigma_{jj'\,m_2\dots m_6} ,
\end{aligned}
\end{equation}
where the index on $v_{ii'}$ is lowered using $g'_{mn}$. Note that
antisymmetrizing the last expression eliminates the $A'$ term. 
We also have, using the fact that $\volG$ is unchanged by a $\E7$
transformation~\cite{csw2}
\begin{equation}
   \ee^{4\Delta'}\det g' = \det g .
\end{equation}
Finally, if one specialises to the case where $U_A{}^B$ parameterises
only the $\SL(8,\bbR)/\SO(8)$ subspace, that is take
$U_{ii'\,jj'}=U^{ii'\,jj'}=0$, we are back to the case discussed in
section~\ref{sec:S-GG}. The three-form $A$ vanishes, and
\begin{equation}
\begin{aligned}
   \dd s^{\prime 2} &= \frac{R^2}{(T^{kl}y_ky_l)^{1/3}}
         \; T^{-1}_{ij} \dd y^i \dd y^j , \\
   \tA' &= -\frac{1}{2(T^{kl}y_ky_l)} \frac{R^6}{5!} \;\epsilon_{i_1\dots i_8}
       (T^{i_1j}y_j)y^{i_2} \dd y^{i_3}\wedge
          \dots \wedge \dd y^{i_8} + \tA , \\
   \ee^{2\Delta'} &= (T^{kl}y_ky_l)^{2/3}
\end{aligned}
\end{equation}
matching the expressions in~\cite{NV,CLP}.


\section{Conclusions}
\label{sec:concl}

This paper presents a unified description of maximally supersymmetric
consistent truncations in terms of generalised geometry. We have seen
that there is a direct analogue of a local group manifold (or
``twisted torus''), namely that the manifold admits what might be
called a \emph{Leibniz generalised parallelisation} (or a
``generalised twisted torus structure''). This means the generalised
tangent space $E$ admits a \emph{global generalised frame}
$\{\hat{E}_A\}$, such that under the generalised Lie derivative
\begin{equation}
   \Lgen_{\hat{E}_A} \hat{E}_B = X_{AB}{}^C \hat{E}_C ,
\end{equation}
with \emph{constant} $X_{AB}{}^C$. In general this defines a
finite-dimensional Leibniz algebra. The existence of such a frame
allows one to consider a supersymmetric generalised Scherk--Schwarz
reduction and $X_{AB}{}^C$ becomes the embedding tensor of the
truncated theory. The key point of this paper was to show that the
``exceptional'' sphere compactifications are actually of this type. This
relied on the demonstration that all round spheres admit Leibniz
generalised parallelisations. Viewed this way, the exceptional sphere
truncations are no different from the conventional Scherk--Schwarz
reductions on a local group manifold. 

A natural question to ask is how the unimodular condition
$f_{ab}{}^b=0$ of conventional Scherk--Schwarz truncations~\cite{ss}
appears in the generalised context. As shown in~\cite{csw2}, and
summarised in  appendix~\ref{app:gen-conn}, the embedding tensor $X$
is equal to the torsion of the generalised Weitzenb\"ock
connection. However generically the torsion is a section of $K\oplus
E^*$~\cite{csw1,csw2} where for $O(d,d)\times\bbR^+$ generalised
geometry $K\simeq\Lambda^3E$ and for $E_{d(d)}\times\bbR^+$
generalised geometry the $K$ representations are listed
in~\cite{csw2}. However, the embedding tensor lies only in the
representation $K$~\cite{embed-lecture} (provided the theory has an
action~\cite{embed-trombone}) so there is a condition that the $E^*$
component vanishes. This is the analogue of the unimodular condition
and reads
\begin{equation}
\label{eq:mod-cond}
   X_{BA}{}^B = 0 .
\end{equation}
In appendix~\ref{app:gen-conn} we calculate $X_{BA}{}^B$ for both the
$O(d,d)\times\bbR^+$~\eqref{eq:uni-Odd} and 
$\Edd\times\bbR^+$~\eqref{eq:uni-Edd} cases, given a conventional
Scherk--Schwarz reduction with flux. If the dilaton $\phi$ and warp
factor $\Delta$ are to be single-valued and bounded we see that 
\begin{equation}
   X_{BA}{}^B = 0 \qquad \Leftrightarrow \qquad
   \begin{cases}
      f_{ab}{}^b=0, \quad \phi=\text{const} & 
         \text{for $O(d,d)\times\bbR^+$}, \\ 
      f_{ab}{}^b=0, \quad \Delta=\text{const} & 
         \text{for $\Edd\times\bbR^+$}.
   \end{cases}
\end{equation}
Thus we indeed reproduce the standard unimodular condition for
Scherk--Schwarz reductions. Note that~\eqref{eq:mod-cond} is also
identically satisfied for the sphere truncations. 

In this paper, we have not proven that the truncations are consistent
but only identified the gauge structure in terms of the frame algebra
and also the scalar field ansatz. In general, one needs ansatze for
the gauge fields and any other tensor fields, as well as the
fermions. In the type IIB case, one already knows~\cite{CLPST1}, for
instance, that consistency requires the correct self-duality condition
on the five-form flux, which we have not considered here. However,
just as in conventional Scherk-Schwarz compactifications, all these
ansatze should follow simply from the existence of a global generalised
frame. For example, the gauge fields appear as sections of $E$ with gauge
transformations generated by the generalised Lie derivative, so
that~\cite{gen-weitz1,EFT1} 
\begin{equation}
   A_\mu = A_\mu^A \hat{E}_A , \qquad
   \delta A_\mu = \der_\mu \Lambda - \Lgen_{A_\mu} \Lambda
       = \left( \der_\mu \Lambda^C
          - X_{AB}{}^C A_\mu^A \Lambda^B \right) \hat{E}_C .
\end{equation}
In general~\cite{EFT1,EFT2}, this will extend to a whole tensor
hierarchy~\cite{tensor-hier}. 

Recall that consistency of a conventional Scherk--Schwarz truncation
is essentially trivial (see for example~\cite{CGLP}). By expanding in
the full set of left-invariant objects one can never generate
something outside the truncation, since such an object is by
definition not left-invariant. Assuming the full ten- or
eleven-dimensional theory can be reformulated with the generalised
geometry manifest, along the lines first suggested in~\cite{dWN},
developed in~\cite{csw2,csw3} for the internal space, and most
recently formulated in full in~\cite{EFT1,EFT2}, one might expect that
the proof of consistency for Leibniz generalised parallelisations is
then equally straightforward.  

Though not the main point of this discussion, it is interesting to
connect the generalised geometry of \cite{csw2} to the internal
``generalised vielbein postulate'' (GVP) of~\cite{dWN,NP,N-new},
used to reformulate eleven-dimensional supergravity in a $4+7$ split
and defined prior to any reference to a  consistent truncation. The
GVP is a differential condition on the generalised split frame
$\{\hat{E}_A\}$~\eqref{eq:E7-split} that has a form reminiscent of the
usual relation between frame and coordinate expressions for a connection,
namely $\der_m\hat{e}^n_a+\Gamma^n{}_{mp}\hat{e}^p_a
=\omega_m{}^b{}_a\hat{e}^n_b$. Originally it was defined for only the
vector component of the frame $\hat{E}^m_A$~\cite{dWN,NP}. This was
then extended to all components in~\cite{N-new}. The generalised
geometry of~\cite{csw2,csw3} gives a precise interpretation of the
GVP. The GVP takes the form 
\begin{equation}
   \nabla_m \hat{E}^M_A + \Xi_{m\ph{M}N}^{\ph{m}M} \hat{E}^N_A
       = \Omega_m{}^B{}_A \hat{E}^M_B , 
\end{equation}
where $\Omega_m$ includes both $\mathcal{Q}_m$ and $\mathcal{P}_m$
defined in~\cite{N-new}, $\Xi$ has a restricted ``triangular'' form
so that, for example $\Xi_{m\ph{m}P}^{\ph{m}n}=0$, and $\nabla$ is the
Levi--Civita connection. This structure matches~\eqref{eq:D-def}
and~\eqref{eq:gen-sc} and so defines a particular generalised
connection $\DGVP_M$. The key point is that only the first component
$\DGVP_m$ is non-zero, where generically, decomposing under
$\GL(7,\bbR)$, we have $\Dgen_M=(\Dgen_m,\Dgen^{m_1m_2},
\Dgen^{m_1m_2\dots m_5}, \Dgen^{m,m_1\dots m_7})$. In fact, we can
identify $\DGVP_M$ directly: it is the standard lift~$\Dgen^\nabla_M$
of the Levi--Civita connection $\nabla$ modified by flux-dependent
terms. Explicitly, using the notation of~\cite{csw2}, we
have\footnote{To write the shift $\Sigma_m$ in terms matching the GVP
   one uses the standard transformations between $\SL(8,\bbR)$ and
   $\SU(8)$ indices as for example in eq.~(B.30) of~\cite{PW}.}
\begin{equation}
   \DGVP_M V^A = \Dgen^\nabla_M V^A + \Sigma_{M\ph{A}B}^{\ph{M}A} V^B
\end{equation}
with
\begin{equation}
\begin{aligned}
   \Sigma_m &= \tfrac{7}{2}\alpha (\der_m\Delta) , & && &&
   \Sigma_{m\,a_1a_2a_3} &= \beta F_{ma_1a_2a_3} , \\
   \Sigma_{m\ph{a}b}^{\ph{m}a} &= \alpha (\der_m\Delta)\delta^a{}_b , 
       & && && 
   \Sigma_{m\,a_1\dots a_6} &= \gamma \tF_{ma_1\dots a_6} ,
\end{aligned}
\end{equation}
Note that $\DGVP_M$ is not torsion-free. It is also an $\E7$-valued
generalised connection, and, as such, in cannot directly act on
spinors to define, for example, the supersymmetry variations. Instead
in~\cite{dWN} the connection is split into $\SU(8)$ irreducible 
representations, and it is shown that the supersymmetry variations can
be written in terms of these pieces, hence fixing the coefficients
$\alpha$, $\beta$ and $\gamma$. As noted, $\DGVP_M$ is defined using
the $\GL(7,\bbR)$ decomposition of $\E7\times\bbR^+$ since, for 
example, only $\DGVP_m$ is non-zero, and $\Xi_m$ has a particular
form. In~\cite{csw2,csw3} a different generalised connection was
defined, that is, in some ways, more natural. By allowing all
components of $\Dgen_M$ to be non-trivial, one can define a direct
analogue of the Levi--Civita connection, namely a torsion-free,
$\SU(8)$ generalised connection, with no need for a decomposition
under $\GL(7,\bbR)$. The internal supersymmetry transformations, and
the bosonic and fermionic equations of motion are then written
directly using $\Dgen_M$. Whether the appearance of this connection
also extends to the full reformulation of the theory, including
dependence on the additional four dimensions, is an interesting open
question. At least for the full $O(10,10)\times\bbR^+$ formulation of
type II theories, we know that the analogous torsion-free connection is
indeed the appropriate object~\cite{siegel,csw1}. 

As a final point, it is obviously of importance to classify what
spaces $M$ admit suitable generalised parallelisations. This would
give a (possibly exhaustive) class of maximal gauged supergravities that
appear as consistent truncations. Note first that the conditions of
generalised parallelisability in general, and the existence of a
Leibniz generalised parallelisation in particular, are much weaker than
the conventional conditions of parallelisability and a local group
manifold structure respectively, as is seen by the $S^5$ and $S^7$
examples. One condition~\cite{GMPW} that can immediately
be derived from the existence of a Leibniz generalised parallelisation
is that $M$ is necessarily a coset space $M=G/H$, where
$\mathfrak{g}$, the Lie algebra of $G$, is a subalgebra of the Leibniz 
algebra~\eqref{eq:embed-def}. A general classification would thus
address the old question of exactly which coset spaces admit
consistent truncations~\cite{CGLP}. Of particular interest is whether
or not the recently discovered family of four-dimensional,
$\mathcal{N}=8$, $\SO(8)$ gaugings~\cite{new-gauge} appear as
truncations within this class.


\acknowledgments

We would like to thank Mariana Gra\~{n}a, Michela Petrini and Chris
Hull for helpful discussions. We especially thank Anthony Ashmore for
permission to use some of his unpublished results on type IIB
$E_{6(6)}\times\bbR^+$ generalised geometry. K.L. is supported by the
National Research Foundation of Korea (NRF) grant funded by the Korea
government (MSIP) with the Grant No. 2005-0049409 (CQUeST).
C.~S-C.~is supported by the German Science Foundation (DFG) under the
Collaborative Research Center (SFB) 676 ``Particles, Strings and the
Early Universe''. D.~W.~is supported by the STFC grant ST/J000353/1
and the EPSRC Programme Grant EP/K034456/1 ``New Geometric Structures
from String Theory''.


\appendix


\section{The round sphere  $S^d$}
\label{app:sphere}

Consider Cartesian coordinates $x^i=ry^i$ with
$\delta_{ij}y^iy^j=1$. The round metric $g$ on $S^d$ of radius $r=R$
is given by
\begin{equation}
   \dd s^2 = R^2 \delta_{ij} \dd y^i \dd y^j .
\end{equation}
One has
\begin{equation}
   \frac{\der}{\der x^i} = y_i \frac{\der}{\der r} + \frac{k_i}{r} ,
\end{equation}
where $k_i$ are the conformal Killing vectors satisfying
\begin{equation}
   \mathcal{L}_{k^i} g = - 2 y^i g .
\end{equation}
In addition one has
\begin{equation}
   k_i(y_j) = i_{k_i}\dd y_j =
   R^{-2} g(k_i,k_j) = R^2 g^{-1}(\dd y_i,\dd y_j)
      = \delta_{ij} - y_iy_j .
\end{equation}
By considering $\tfrac{1}{(d+1)!}\epsilon_{i_1\dots i_{d+1}}\dd
x^{i_1}\wedge \dd x^{i_{d+1}}$ one can write the volume form on $S^d$
as 
\begin{equation}
\label{eq:vol}
   \vol_g = \frac{R^d}{d!}\epsilon_{i_1\dots i_{d+1}}
          y^{i_1}\dd y^{i_2}\wedge \dots \wedge \dd y^{i_{d+1}} .
\end{equation}
We define the $\SO(d+1)$ Killing vectors
\begin{equation}
   v_{ij} = R^{-1}\left( y_ik_j - y_jk_i \right) ,
\end{equation}
such that under the Lie bracket
\begin{equation}
\begin{aligned}
   \BLie{v_{ij}}{v_{kl}}
      &= R^{-1}\left(\delta_{ik}v_{lj} - \delta_{il}v_{kj}
          - \delta_{jk}v_{li} + \delta_{jl}v_{ki} \right) , \\
   \mathcal{L}_{v_{ij}}y_k
      &= R^{-1} \left( y_i\delta_{jk} - y_j\delta_{ik} \right) , \\
   \mathcal{L}_{v_{ij}}\dd y_k
      &= R^{-1} \left( \dd y_i\delta_{jk} - \dd y_j\delta_{ik} \right) .
\end{aligned}
\end{equation}

It is also useful to define
\begin{equation}
\label{eq:ost-def}
\begin{aligned}
   \omega_{ij} &= R^2 \dd y_i \wedge \dd y_j , \\
   \sigma_{ij} = {}* \omega_{ij}
      &= \frac{R^{d-2}}{(d-2)!}\epsilon_{ijk_1\dots k_{d-1}}
          y^{k_1}\dd y^{k_2}\wedge \dd y^{k_{d-1}} , \\
   \tau_{ij} &= R (y_i\dd y_j - y_j\dd y_i)\otimes \vol_g
\end{aligned}
\end{equation}
Since $y^i$ and $\dd y^i$ transform in the fundamental
representation under $\mathcal{L}_{v_{ij}}$ and $\vol_g$ is invariant,
we immediately have that all these tensors transform in the adjoint
representation, that is,
\begin{equation}
\label{eq:Lie-ders}
\begin{aligned}
   \mathcal{L}_{v_{ij}}\omega_{kl}
      &= R^{-1}\left(\delta_{ik}\omega_{lj}
          - \delta_{il}\omega_{kj}
          - \delta_{jk}\omega_{li}
          + \delta_{jl}\omega_{ki} \right) , \\
   \mathcal{L}_{v_{ij}}\sigma_{kl}
      &= R^{-1}\left(\delta_{ik}\sigma_{lj}
          - \delta_{il}\sigma_{kj}
          - \delta_{jk}\sigma_{li}
          + \delta_{jl}\sigma_{ki} \right) , \\
   \mathcal{L}_{v_{ij}}\tau_{kl}
      &= R^{-1}\left(\delta_{ik}\tau_{lj}
          - \delta_{il}\tau_{kj}
          - \delta_{jk}\tau_{li}
          + \delta_{jl}\tau_{ki} \right) .
\end{aligned}
\end{equation}
We also have, contracting indices with the sphere metric,
\begin{equation}
\label{eq:contractions}
\begin{aligned}
   v_{ij}\cdot v_{kl} &:= (v_{ij})^m(v_{kl})_m \\
      &= y_iy_k\delta_{jl} - y_jy_k\delta_{il}
          - y_iy_l\delta_{jk} + y_jy_l\delta_{ik} , \\
   \omega_{ij}\cdot \omega_{kl}
      &:= \tfrac{1}{2}(\omega_{ij})^{mn}(\omega_{kl})_{mn} \\
      &= \delta_{ik}\delta_{jl} - \delta_{il}\delta_{jk}
          - \left( y_iy_k\delta_{jl} - y_jy_k\delta_{il}
             - y_iy_l\delta_{jk} + y_jy_l\delta_{ik} \right), \\
   \sigma_{ij}\cdot \sigma_{kl}
      &:= \tfrac{1}{(d-2)!}(\sigma_{ij})^{m_1\dots m_{d-2}}
           (\sigma_{kl})_{m_1\dots m_{d-2}} \\
      &= \delta_{ik}\delta_{jl} - \delta_{il}\delta_{jk}
          - \left( y_iy_k\delta_{jl} - y_jy_k\delta_{il}
             - y_iy_l\delta_{jk} + y_jy_l\delta_{ik} \right), \\
   \tau_{ij}\cdot \tau_{kl}
      &:= \tfrac{1}{d!}(\tau_{ij})^{m,n_1\dots n_d}
          (\tau_{kl})_{m,n_1\dots n_d} \\
      &= y_iy_k\delta_{jl} - y_jy_k\delta_{il}
          - y_iy_l\delta_{jk} + y_jy_l\delta_{ik} .
\end{aligned}
\end{equation}
Finally we note
\begin{equation}
\label{eq:vol-identity}
\begin{aligned}
   i_{v_{ij}}\vol_g
     &= - \frac{R^{d-1}}{(d-1)!} \left(
          y_i\epsilon_{jk_1\dots k_d}-y_j
             \epsilon_{ik_1\dots k_d}\right)
             y^{k_1}\dd y^{k_2}\wedge\dots\wedge\dd y^{k_d} \\
     &= \frac{(d-1)R^{d-1}}{(d-1)!}
          y_{k_1}\epsilon_{ijk_2\dots k_d}
             y^{[k_1}\dd y^{k_2}\wedge\dots\wedge\dd y^{k_d]} \\
     &= \frac{R^{d-1}}{(d-1)!}
          \epsilon_{ijk_1\dots k_{d-1}}
          \dd y^{k_1}\wedge\dots\wedge\dd y^{k_{d-1}} \\
     &=  \frac{R}{d-1} \dd\sigma_{ij}
\end{aligned}
\end{equation}
where in the going to the second line we use
$y_{[i_1}\epsilon_{i_2\dots i_{d+2}]}=0$.

We also use a couple of further identities. Defining the
set of tensors $A_i{}^m{}_n=v^m_{ij}\der_n y^j$ we have
\begin{equation}
   A_i{}^m{}_n \der_m y^k = \left( i_{v_{ij}}\dd y^k \right) \der_n y^j
      = R^{-1}\left( y_i\delta^k_j - y_j\delta^k_i \right) \der_n y^j
      = R^{-1}y_i \der_n y^k .
\end{equation}
Since $\dd y^i$ is an overcomplete basis for $T^*M$ this implies that
$A_i$ is proportional to the identity matrix, namely,
\begin{equation}
\label{eq:metric-id}
   v^m_{ij}\der_n y^j = \left(R^{-1}y_i\right) \delta^m_n .
\end{equation}
Finally, suppose we have a metric
\begin{equation}
   \dd s^{\prime 2} = R^2 T^{-1}_{ij} \dd y^i\dd y^j ,
\end{equation}
then~\cite{NVvN,NV} one has
\begin{equation}
\label{eq:det-id}
   \det g' = \frac{(T_{ij}y^iy^j)}{\det T} \det g .
\end{equation}
This can be seen by considering the variation with respect to $T_{ij}$
\begin{equation}
   \delta \log \det g' = {g'}^{mn} \delta {g'}_{mn} = R^2 \delta
      (T^{-1}_{ij}) \, {g'}^{-1} (\dd y^i, \dd y^j) . 
\end{equation}
Using~\eqref{eq:metric-id}, one has
\begin{equation}
   {g'}^{mn} = \tfrac12 (T^{ij} y_i y_j){}^{-1} T^{kl} T^{k'l'}
      (v_{kk'}){}^m (v_{ll'})^n , 
\end{equation}
leading to
\begin{equation}
   \delta \log \det g' = \delta \log \frac{(T_{ij}y^iy^j)}{\det T} , 
\end{equation}
which integrates to~\eqref{eq:det-id}.


\section{Type IIB $E_{6(6)}$ generalised geometry}
\label{app:e6-gen-geom}


In this appendix we summarise the main ingredients of
$E_{6(6)}\times\bbR^+$ generalised geometry as applied to type IIB
supergravity. The form of the generalised tangent space was first
given in~\cite{chris}. The patching, generalised Lie derivative, and
form of the split frame are given implicitly in~\cite{csw2} after
applying the IIB decomposition described in the appendix~C of that
paper. Several of the explicit expressions given here were derived in
unpublished work by Ashmore~\cite{anthony} and we are very grateful
for the permission to summarise them.

One considers the 27-dimensional generalised tangent space~\cite{chris}
\begin{equation}
\label{eq:E-e6}
\begin{aligned}
   E &\simeq TM \oplus (T^*M \oplus T^*M) \oplus \Lambda^3T^*M
      \oplus (\Lambda^5T^*M \oplus \Lambda^5T^*M) , \\
   V &= v + \rho^\xa + \lambda + \chi^\xa .
\end{aligned}
\end{equation}
This transforms in the $\rep{27}_{\rep{1}}$ representation of
$E_{6(6)}\times\bbR^+$, with a weight one under the $\bbR^+$-factor,
where a scalar $\rep{1}_{\rep{k}}$ of weight $k$ is a section of
$(\det T^*M)^{k/3}$~\cite{csw2}. The split of $V$ above represents the
decomposition under a $\SL(2,\bbR)\times\GL(5,\bbR)$ subgroup where
$\SL(2,\bbR)$ is the type IIB S-duality group. The symmetric
$E_{6(6)}$ cubic invariant is given by~\cite{anthony}
\begin{equation}
   c(V,V,V) = \tfrac{1}{2}i_v\lambda \wedge \lambda
       + \tfrac{1}{2} \lambda\wedge\rho_\xa\wedge\rho^\xa
       +  (i_v\rho_\xa) \chi^\xa ,
\end{equation}
where we lower $\SL(2,\bbR)$ indices by
$u_\xa=\epsilon_{\xa\xb}u^\xb$. Note that this is a five-form because
of the weight of the generalised vector. There is a nilpotent subgroup of
$E_{6(6)}$ that acts as~\cite{anthony}
\begin{equation}
\begin{aligned}
   \ee^{B^\xa+A} V
     &= v - i_v B^\xa - i_v A
        - \tfrac{1}{2}B_\xa\wedge i_v B^\xa
        - B^\xa \wedge i_v A
        - \tfrac{1}{6}B^\xa\wedge B_\xb\wedge i_vB^\xb
        \\ & \qquad
        + \rho^\xa + B_\xa\wedge \rho^\xa
        - A \wedge \rho^\xa
        + \tfrac{1}{2}B^\xa \wedge B_\xb\wedge \rho^\xb
        \\ & \qquad
        + \lambda + B^\xa\wedge\lambda
        + \chi^\xa ,
\end{aligned}
\end{equation}
where $B^\xa\in\Lambda^2T^*M$ and $A\in\Lambda^4T^*M$. As
in~\eqref{eq:patch} the generalised tangent space is really patched by
\begin{equation}
   V_{(i)} = \ee^{\dd\hat{\Lambda}_{(ij)}^\xa+\dd\Lambda_{(ij)}}
        V_{(j)} .
\end{equation}
If $B^\xa$ and $A$ are two-form and four-form gauge potentials patched
by
\begin{equation}
\begin{aligned}
   B^\xa_{(i)} &= B^\xa_{(j)} + \dd\hat{\Lambda}^\xa_{(ij)} , \\
   A_{(i)} &= A_{(j)} + \dd\Lambda_{(ij)}
      + \tfrac{1}{2} \dd\hat{\Lambda}_{(ij)\,\xa}
         \wedge B^\xa_{(j)} , 
\end{aligned}
\end{equation}
the corresponding gauge-invariant field strengths are
\begin{equation}
\label{eq:IIB-flux}
   H^\xa = \dd B^\xa , \qquad
   F = \dd A - \tfrac{1}{2}B_\xa\wedge \dd B^\xa .
\end{equation}
As in~\eqref{eq:iso} we can use the gauge potentials to define the
isomorphism in~\eqref{eq:E-e6} by
\begin{equation}
   V = \ee^{B^\xa+A}\tilde{V} ,
\end{equation}
where $\tilde{V}$ is a sum a vector and $p$-forms (without additional
patching). Given a pair of generalised vectors we have the generalised
Lie derivative~\cite{csw2,anthony}
\begin{equation}
\label{eq:Lgen-e6}
\begin{aligned}
   \Lgen_V V' &= (V\cdot\der)V' - (\der \oadj V)V' \\
      &= \BLie{v}{v'}
      + \mathcal{L}_v\rho^{\prime \xa}
      - i_{v'}\dd \rho^\xa
      + \mathcal{L}_v\lambda - i_{v'}\dd\lambda
          + \dd\rho_\xa\wedge\rho^{\prime \xa} \\ & \qquad
      + \mathcal{L}_v\chi^{\prime\xa}
          - \dd\lambda\wedge\rho^{\prime\xa}
          + \dd\rho^{\xa}\wedge\lambda' ,
\end{aligned}
\end{equation}
where $\oadj$ projects onto the $E_{6(6)}\times\bbR^+$ adjoint.

Let $\hat{f}_\za{}^\xa$ be an $\SL(2,\bbR)$ frame, and $f^\za{}_\xa$
the dual frame, which we can write explicitly in terms a
parametrisation of $\SL(2,\bbR)/\SO(2)$ as 
\begin{equation}
\label{eq:SL2-frame}
   \hat{f}_\za{}^\xa
       = \begin{pmatrix}
             \ee^{\phi/2} & C \ee^{\phi/2} \\  0 & \ee^{-\phi/2}
          \end{pmatrix} , \qquad
   f^\za{}_\xa
       = \begin{pmatrix}
             \ee^{-\phi/2} & 0 \\ - C \ee^{\phi/2} & \ee^{\phi/2}
          \end{pmatrix} .
\end{equation}
If $\hat{e}_a$ and $e^a$ are a conventional frame for $TM$ and its
dual, then we can define a split frame by~\cite{csw2,anthony}
\begin{equation}
\label{eq:E6-split}
\begin{aligned}
   \hat{E}_a
      &= \ee^\Delta\big(
         \hat{e}_a - i_{\hat{e}_a} B^\xa - i_{\hat{e}_a} A
         - \tfrac{1}{2}B_\xa\wedge i_{\hat{e}_a} B^\xa
         \\ & \qquad \qquad \qquad
         - B^\xa \wedge i_{\hat{e}_a} A
         - \tfrac{1}{6}B^\xa\wedge B_\xb\wedge i_{\hat{e}_a}B^\xb
         \big), \\
   \hat{E}^a_\za
      &= \ee^\Delta\ee^{-\phi/2}\big(
         \hat{f}_\za{}^\xa e^a + B_\za\wedge e^a
         - \hat{f}_\za{}^\xa A \wedge e^a
         + \tfrac{1}{2}B^\xa \wedge B_\za\wedge e^a \big) ,
         \\
   \hat{E}^{abc}
      &= \ee^\Delta\ee^{-\phi} \left(
         e^{abc} + B^\xa\wedge e^{abc}\right) ,
         \\
   \hat{E}^{a_1\dots a_5}_\za
      &= \ee^\Delta\ee^{-3\phi/2} \hat{f}_\za{}^\xa e^{a_1\dots a_5} , 
\end{aligned}
\end{equation}
where $B_\za=\hat{f}_\za{}^\xa
B_\xa=\hat{f}_\za{}^\xa\epsilon_{\xa\xb}B^\xb=\epsilon_{\za\zb}f^\zb{}_\xa
B^\xa$ and $e^{a_1\dots a_n} =e^{a_1}\wedge\dots\wedge e^{a_n}$. Here
the choice of powers of the dilaton means that $\hat{e}_a$ are
vielbeins for a string-frame metric. The warp-factor $\Delta$ is
associated to compactifications with a string-frame metric of the form
\begin{equation}
   \dd s^2 = \ee^{2\Delta} \dd s^2_{1,4} + \dd s^2(M)
\end{equation}
where $\dd s^2_{1,4}$ is the metric in the non-compact five-dimensional
space. Note that with the $\SL(2,\bbR)$ frame~\eqref{eq:SL2-frame} we
can define the complex three-form field strength charged under $U(1)\simeq\SO(2)$
\begin{equation}
\begin{aligned}
   G &= - \big( H^{\hat{1}}+\ii H^{\hat{2}} \big)
     = - \ee^{-\phi/2}\dd B^1
          - \ii \ee^{\phi/2}\left( \dd B^2 - \chi \dd B^1\right) \\
     & = \ii\, \ee^{\phi/2}\left( \tau \dd B^1 - \dd B^2 \right) ,
\end{aligned}
\end{equation}
where $\tau=C+\ii\ee^{-\phi}$. We then have the Bianchi identity
for the five-form field strength~\eqref{eq:IIB-flux}
\begin{equation}
   \dd F = - \tfrac{1}{2} H_\xa \wedge H^\xa
      = H^1\wedge H^2
      = \tfrac{1}{2}\ii G \wedge G^* .
\end{equation}
We see that our conventions for the gauge potentials and axion and
dilaton match the standard definitions, as for example
in~\cite{IIB-conv}. The NSNS two-form is $B^1$, while the RR
potentials are $C_{(0)}=C$, $C_{(2)}=B^2$ and $C_{(4)}=A$.

Using the split frame we can define the generalised
metric\footnote{This is not to be confused with the complex three-form
   $G$ just defined.} $G$ by~\cite{csw2}
\begin{equation}
   G(\hat{E}_A,\hat{E}_B) = \delta_{AB}
\end{equation}
where given $\{\hat{E}_A\}=\{\hat{E}_a,\hat{E}^a_\za,
\hat{E}^{abc},\hat{E}^{a_1\dots a_5}_\za\}$ we define (compatible with
the conventions mentioned in footnote~\ref{sum-conv})
\begin{equation}
\begin{aligned}
   \delta_{a,b} &= \delta_{ab} , & && &&
   \delta^{a_1a_2a_3,b_1b_2b_3}
      &= {3!}\delta^{[a_1|b_1|}\delta^{a_2|b_2|}\delta^{a_3]b_3} \\
   \delta^{a,b}_{\za,\zb} &=  \delta_{\za\zb} \delta^{ab} & && &&
   \delta^{a_1\dots a_5,}_\za{}^{b_1\dots b_5}_\zb
      &= {5!}\delta_{\za\zb}\delta^{[a_1|b_1|}\delta^{a_2|b_2|}
            \dots\delta^{a_5]b_5} ,
\end{aligned}
\end{equation}
with all other components vanishing. Equivalently we can define the
inverse generalised metric as
\begin{equation}
   G^{-1\, MN} = \delta^{AB}\hat{E}_A^M\hat{E}_B^N .
\end{equation}
In components, we note in particular that
\begin{equation}
\label{eq:G-comp}
\begin{aligned}
   G^{-1\, m,n} &= \ee^{2\Delta} g^{mn} , \\
   G^{-1}{}^{m,}{}^\xb_n &= \ee^{2\Delta} B^{\xb\, m}{}_n , \\
   G^{-1}{}^{m,}{}^\xb_{n_1n_2n_3}
       &= - \ee^{2\Delta} \left( A^m{}_{n_1n_2n_3}
          - \tfrac{3}{2}B_{\xa [n_1n_2}B^{\xa\, m}{}_{n_3]} \right) .
\end{aligned}
\end{equation}
and
\begin{equation}
\label{eq:G-comp2}
   G^{-1\, \xa,\xb}_{\ph{-1\,}m,n}
      = \ee^{2\Delta}\left(
         \ee^{-\phi} h^{\xa\xb}g_{mn} - B^\xa_{mp}g^{pq}B^\xb_{qn}
         \right) ,
\end{equation}
where $h^{\xa\xb}=\delta^{\za\zb}\hat{f}_\za{}^\xa\hat{f}_\zb{}^\xb$
is the inverse $\SL(2,\bbR)$ metric. Explicity one has
\begin{equation}
   \ee^{-\phi} h^{\xa \xb}
       = \begin{pmatrix}
             1 && C \\  C && C^2 + \ee^{-2\phi}
          \end{pmatrix}
\end{equation}
%


\section{Generalised connections and conventional 
   Scherk--Schwarz} 
\label{app:gen-conn}


In this appendix we recall and expand slightly two of the results
of~\cite{csw2}. First is the relationship between the embedding tensor
and the torsion of the generalised Weitzenb\"ock connection. Second is
the calculation of the embedding tensor for the specific example of a
conventional Scherk--Schwarz reduction on a local group manifold
$M$. In~\cite{csw2} the calculation was for $\Edd\times\bbR^+$
generalised geometry. Here we also consider the $O(d,d)\times\bbR^+$
case. 

Recall that, given a conventional parallelisation, there is a unique
connection $\hat{\nabla}_m$, known as the Weitzenb\"ock connection,
that preserves the frame, that is
$\hat{\nabla}_m\hat{e}^n_a=0$. However, generically $\hat{\nabla}_m$
is not torsion-free, instead, the torsion $T^m{}_{np}$ is related to
the Lie algebra structure constants,
\begin{equation}
\label{eq:conv-wc}
   T^c{}_{ab} = - f_{ab}{}^c , \qquad
   \BLie{\hat{e}_a}{\hat{e}_b} = f_{ab}{}^c \hat{e}_c .
\end{equation}
Let us now see how the analogous concepts arise in generalised
geometry. 

A generalised connection~\cite{g-conn,csw1,csw2} is a first-order
linear differential operator $D_M$ which acts on generalised vectors
as 
\begin{equation}
\label{eq:D-def}
   D_M V^N = \der_M V^N + \Gamma_{M\ph{N}P}^{\ph{M}N} V^P .
\end{equation}
Acting on a local frame $\{\hat{E}_A\}$ one can define the analogue of
the spin connection
\begin{equation}
\label{eq:gen-sc}
   D_M \hat{E}^N_A = \Omega_{M\ph{B}A}^{\ph{M}B} \hat{E}_B^N .
\end{equation}
The generalised one-forms $\Omega^A_{\ph{A}B}$ are Lie-algebra
valued. If the corresponding group is $H$ we have an $H$-compatible
generalised connection. If $H\subseteq G$, where $G$ is the
generalised structure group $G$ (here $\Edd\times\bbR^+$ or
$O(d,d)\times\bbR^+$), we can also always define the torsion $T$ of
the generalised connection as~\cite{csw1,csw2}\footnote{Note that for
   $O(d,d)\times\bbR^+$ connections we are taking a slightly different
   convention from~\cite{csw1} for the ordering of the indices in
   $T$, so as to give a uniform treatment with the $\Edd\times\bbR^+$
   case.}, given $V\in E$,
\begin{equation}
\label{eq:T-def}
   T(V) = \Lgen_V^D - \Lgen_V
\end{equation}
where $T(V)^N{}_P=V^MT_M{}^N{}_P$ is an element of the adjoint
representation of $G$. Note that in general the torsion lies in only
particular irreducible representations of $G$~\cite{csw1,csw2}
\begin{equation}
   T\in K\oplus E^* , 
\end{equation}
where for $O(d,d)\times\bbR^+$ we have  $E\simeq E^*$ and
$K=\Lambda^3E$, while for $\Edd\times\bbR^+$ one finds $K$ transforms
in the same representation as the embedding tensor, for example
$\rep{912}$ for $\E7$ and $\rep{351}$ for $E_{6(6)}$.  The key results 
of~\cite{csw1,csw2} are first that
\begin{quote}
   \emph{There always exists a torsion-free, $H$-compatible
      generalised connection, where $H$ is the maximally compact
      subgroup of $G$.} 
\end{quote}
and second that, although this connection is not unique, there is
a unique Ricci tensor which captures the bosonic equations of 
motion on the compactification space. (For $O(d,d)$ this was first
described using the DFT formalism in~\cite{siegel}
and~\cite{JLP}.) Furthermore the internal contributions to the
supersymmetry variations can be written in terms of unique
$H$-covariant projections of the connection, the generalised geometric
analogues of the Dirac operator~\cite{csw1,csw2,csw3}. 

Just as in the conventional case, $\Omega$ is a global section of
$E^*$ if and only if $\{\hat{E}_A\}$ is globally defined. If this is
the case, given any generalised connection $D$, one can always define
a unique new connection $\hat{D}=D-\Omega$ which satisfies
\begin{equation}
   \hat{\Dgen}_M \hat{E}^N_A = 0 .
\end{equation}
This is the generalised Weitzenb\"ock
connection~\cite{gen-weitz1,gen-weitz2}. As in the conventional case,
the structure constants of the frame algebra are given by the
generalised torsion (in frame indices) of the generalised
Weitzenb\"ock connection~\cite{csw1,csw2}\footnote{Note
   that for $O(d,d)\times\bbR^+$ generalised geometry~\cite{csw1}, to
   incorporate the dilaton and $O(d,d)$ spinors correctly, one
   actually considers a ``weighted'' generalised tangent space 
   $\tilde{E}\simeq(\det T^*M)\otimes(TM\oplus T^*M)$ with a
   ``conformal basis'' $\{\hat{E}_A\}$ (cf.~\eqref{eq:Odd-norm})
   satisfying $\eta(\hat{E}_A,\hat{E}_B)=\Phi^2 \eta_{AB}$ where
   $\Phi\in\det T^*M$. The generalised torsion of the corresponding
   Weitzenb\"ock connection is then actually given by  
   \begin{equation*}
      X_{AB}{}^C = E^C \cdot \big(\Lgen_{\Phi^{-1}\hat{E}_A} \hat{E}_B \big)
         = - T_A{}^C{}_B . 
   \end{equation*}
   }
\begin{equation}
   X_{AB}{}^C = E^C \cdot \big(\Lgen_{\hat{E}_A} \hat{E}_B \big)
      = - T_A{}^C{}_B ,
\end{equation}
where $\{E^A\}$ is the dual generalised basis on $E^*$.

Now suppose the generalised parallelisation arises from a
conventional local-group manifold. Let $\hat{e}_a$ be an invariant
global frame for $TM$, for example the left-invariant vector
fields. Let $e^a$ be the dual frame for $T^*M$. The split
frame~\eqref{eq:Odd-split} for $O(d,d)$ or~\eqref{eq:E7-split}
for~$\E7$ (more generally see~eq.~(3.19) of~\cite{csw1} and eq.~(2.15)
of \cite{csw2}) is globally defined, and gives a generalised 
parallelisation. Furthermore, we can identify the generalised
Weitzenb\"ock connection as the lift $\Dgen^{\hat{\nabla}}_M$, as
defined in~\cite{csw1,csw2}, of the conventional Weitzenb\"ock
connection $\hat{\nabla}_m$. The corresponding torsion was calculated
in~\cite{csw1,csw2}. One finds, for $O(d,d)\times\bbR^+$, that the 
non-vanishing elements of the frame algebra are 
\begin{equation}
\begin{aligned} 
   \Lgen_{\Phi^{-1}\hat{E}_a} \hat{E}_b
      &= f_{ab}{}^c \hat{E}_c + H_{abc} \hat{E}^c 
           - (f_{ac}{}^c + 2\der_a\phi) \hat{E}_b , \\
   \Lgen_{\Phi^{-1}\hat{E}_a} \hat{E}^b
      &= - f_{ac}{}^b \hat{E}^c 
           - (f_{ac}{}^c + 2\der_a\phi) \hat{E}^b , \\
   \Lgen_{\Phi^{-1}\hat{E}^a} \hat{E}_b
      &= f_{bc}{}^a \hat{E}^c , 
\end{aligned}
\end{equation}
where $H_{abc}$ and $\der_a\phi$ are the frame components of the flux
and the derivative of the dilaton. For $\Edd\times\bbR^+$ the
non-vanishing elements are 
\begin{align*}
   \Lgen_{\hat{E}_a} \hat{E}_b &= \ee^\Delta \Big[ 
      f_{ab}{}^c\hat{E}_c 
       + \tfrac{1}{2!} F_{abc_1c_2} \hat{E}^{c_1c_2}
       + \tfrac{1}{5!} \tF_{abc_1\dots c_5} \hat{E}^{c_1\dots c_5}
       \\* & \qquad \qquad 
       + (\der_a \Delta) \hat{E}_b 
       - (\der_b \Delta) \hat{E}_a  \Big] , \\
   \Lgen_{\hat{E}_a} \hat{E}^{b_1b_2} &= \ee^\Delta \Big[ 
       - 2f_{ac}{}^{[b_1} \hat{E}^{|c|b_2]}
       + \tfrac{1}{3!} F_{a c_1 \dots c_3} \hat{E}^{b_1b_2c_1 \dots c_3}
       \\* & \qquad \qquad
       - \tfrac{1}{5!} \tF_{acc_1 \dots c_5}\hat{E}^{c,b_1b_2c_1\dots c_5}
       + (\der_a \Delta) \hat{E}^{b_1b_2} 
       + 2 (\der_c \Delta) \delta_a{}^{[b_1} \hat{E}^{|c| b_2]} \Big] , \\
   \Lgen_{\hat{E}_a} \hat{E}^{b_1\dots b_5}  &= \ee^\Delta \Big[ 
       - 5f_{ac}{}^{[b_1} \hat{E}^{|c|b_2b_3b_4b_5]}
       + \tfrac{1}{2!} F_{acc_1c_2}\hat{E}^{c,b_1\dots b_5c_1c_2}   
       \\* &\qquad \qquad 
       + (\der_a \Delta) \hat{E}^{b_1\dots b_5} 
       + 5 (\der_c \Delta) \delta_a{}^{[b_1} \hat{E}^{|c| b_2 \dots b_5]} 
       \Big] , \\
   \Lgen_{\hat{E}_a} \hat{E}^{b,b_1\dots b_7}&= \ee^\Delta \Big[ 
       - f_{ac}{}^{b} \hat{E}^{c,b_1\dots b_7}
       - 7f_{ac}{}^{[b_1} \hat{E}^{|b,c|b_2\dots b_7]}
       \\* & \qquad \qquad 
       + (\der_a \Delta) \hat{E}^{b,b_1\dots b_7} 
       + (\der_c \Delta) \delta_a{}^{b} \hat{E}^{c, b_1 \dots b_7}
       + 7 (\der_c \Delta) \delta_a{}^{[b_1} \hat{E}^{|b, c| b_2 \dots b_7]}
       \Big], \\
   \Lgen_{\hat{E}^{a_1a_2}} \hat{E}_b &=  \ee^\Delta \Big[ 
       2f_{bc}{}^{[a_1} \hat{E}^{|c|a_2]}
       + f_{c_1c_2}{}^{[a_1}\delta^{a_2]}_b\hat{E}^{c_1c_2}
       \\* & \qquad \qquad
       - \tfrac{6}{4!} F_{c_1 \dots c_4}\delta_b^{[a_1}\hat{E}^{a_2c_1 \dots c_4]}   
       - 3(\der_c\Delta) \delta^{[c}_b\hat{E}^{a_1a_2]}\Big]  , \\
   \Lgen_{\hat{E}^{a_1a_2}} \hat{E}^{b_1b_2} &= \ee^\Delta \Big[ 
       f_{c_1c_2}{}^{[a_1} \hat{E}^{a_2]b_1b_2c_1c_2}
       \\* & \qquad \qquad 
       - \tfrac{2}{4!} F_{c_1 \dots c_4}\hat{E}^{[b_1 ,b_2]a_1a_2c_1\dots c_4}
       - (\der_c\Delta) \hat{E}^{ca_1a_2b_1b_2}\Big] , \\
   \Lgen_{\hat{E}^{a_1a_2}} \hat{E}^{b_1 \dots b_5} &= \ee^\Delta \Big[ 
       f_{c_1c_2}{}^{[a_1} \hat{E}^{a_2],b_1\dots b_5c_1c_2}
       + 2 f_{c_1c_2}{}^{[a_1} \hat{E}^{|c_1,c_2|a_2]b_1\dots b_5}
       \\* & \qquad \qquad
       - 5 (\der_c\Delta) \hat{E}^{[b_1,b_2\dots b_5]ca_1a_2} \Big] , \\
   \Lgen_{\hat{E}^{a_1 \dots a_5}} \hat{E}_b &= \ee^\Delta \Big[ 
       5f_{bc}{}^{[a_1} \hat{E}^{|c|a_2\dots a_5]}
       + 10f_{c_1c_2}{}^{[a_1}\delta^{a_2}_b\hat{E}^{a_3a_4a_5]c_1c_2}
       \\* & \qquad \qquad 
       - 6(\der_c\Delta) \delta^{[c}_b\hat{E}^{a_1\dots a_5]} \Big] , \\
   \Lgen_{\hat{E}^{a_1\dots a_5}} \hat{E}^{b_1b_2} &= \ee^\Delta \Big[ 
       - 10 f_{c_1c_2}{}^{[a_1} \hat{E}^{a_2,a_3a_4a_5]b_1b_2c_1c_2}
       - 5 f_{c_1c_2}{}^{[a_1} \hat{E}^{|c_1,c_2|a_2\dots a_5] b_1b_2}
       \\* & \qquad \qquad
       - 2 (\der_c\Delta) \hat{E}^{[b_1,b_2]ca_1\dots a_5}\Big] ,
\end{align*}
where again $F_{abcd}$ and $\tF_{a_1\dots a_7}$ are the frame
components of the fluxes and $\der_a\Delta$ is the frame component of
the derivative of the warp factor. We see that in each case, provided
the frame components of the fluxes and $\der_a\phi$ and $\der_a\Delta$
are constant, then we are indeed in the class of generalised
parallelisations with constant $X_{AB}{}^C$, that is we have a
Leibniz generalised parallelisation. If we take
$\der_a\phi=f_{ab}{}^a=0$ or $\der_a\Delta=f_{ab}{}^a=0$ we see that
these frame algebras match the standard gaugings in the
literature~\cite{KM,HR-E1,GMPW} and~\cite{dAFT,N-embed}. 

We can also calculate the trace $X_A=X_{BA}{}^B$. We find that, for
the $O(d,d)\times\bbR^+$ case, the only non-zero components are 
\begin{equation}
\label{eq:uni-Odd}
  X_a = - (f_{ab}{}^b + 2\der_a\phi)d , 
\end{equation}
while for $\Edd\times\bbR^+$, they are 
\begin{equation}
\label{eq:uni-Edd}
  X_a = - \left[f_{ab}{}^b - (9-d)\der_a\Delta\right]k , 
\end{equation}
where $k$ is a factor depending on the dimension $d$. (These 
expressions are most easily calculated by considering the generalised
Lie derivative of the volume forms~\cite{csw1,csw2}
$\Phi=\sqrt{g}\ee^{-2\phi}$ and $\volG=\sqrt{g}\ee^{(9-d)\Delta}$
respectively.)




\end{document}